\documentclass[twoside,11pt]{article}

\usepackage[abbrvbib, preprint]{jmlr2e}

\usepackage{pifont}
\usepackage{multicol}
\usepackage{hhline,bm,amssymb,epsfig,amsmath,array,xcolor,hyperref}
\usepackage{graphicx,soul}
\usepackage{multirow}
\usepackage{graphicx}
\usepackage{booktabs,natbib}

\def\argmin{\mathop{\rm argmin}}

\def\Var{\mathop{\rm Var}}
\def\sign{\mathop{\rm sign}}

\newcommand{\bff}{\mathbf{f}}

\newcommand{\xmark}{\ding{55}}%
\newcommand{\cmark}{\ding{51}}%

\newcommand*{\EOP}{\hfill\ensuremath{\square}}%

\jmlrheading{1}{2021}{1-48}{4/00}{10/00}{meila00a}{Ben Dai, Xiaotong Shen, and Wei Pan}

\ShortHeadings{Two-level monotonic multistage recommender systems}{Ben Dai, Xiaotong Shen, and Wei Pan}
\firstpageno{1}

\begin{document}

\title{Two-level monotonic multistage recommender systems\thanks{Research supported in part by 
NSF grants DMS-1712564, DMS-1721216, DMS-1952539, and NIH grants 1R01GM126002, R01HL105397, 
R01AG069895. 
}}


\author{\name Ben Dai$^{1,3}$ and Xiaotong Shen$^1$ and Wei Pan$^2$\\
	$^1$\addr Department of Statistics, University of Minnesota\\
	$^2$\addr Division of Biostatistics, University of Minnesota\\ 
	$^3$\addr Department of Statistics, The Chinese University of Hong Kong\\
	\email bendai@cuhk.edu.hk; xshen@umn.edu; panxx014@umn.edu}


\maketitle

\begin{abstract}
	A recommender system learns to predict the user-specific preference or intention over many items simultaneously for all users, making personalized recommendations based on a relatively small number of observations. One central issue is how to leverage three-way interactions, referred to as user-item-stage dependencies on a monotonic chain of events, to enhance the prediction accuracy. A monotonic chain of events occurs, for instance, in an article sharing dataset, where a ``follow'' action implies a ``like'' action, which in turn implies a ``view'' action. In this article, we develop a multistage recommender system utilizing a two-level monotonic property characterizing a monotonic chain of events for personalized prediction. Particularly, we derive a large-margin classifier based on a nonnegative additive latent factor model in the presence of a high percentage of missing observations, particularly between stages, reducing the number of model parameters for personalized prediction while guaranteeing prediction consistency. On this ground, we derive a regularized cost function to learn user-specific behaviors at different stages, linking decision functions to numerical and categorical covariates to model user-item-stage interactions. Computationally, we derive an algorithm based on blockwise coordinate descent. Theoretically, we show that the two-level monotonic property enhances the accuracy of learning as compared to a standard method treating each stage individually and an ordinal method utilizing only one-level monotonicity. Finally, the proposed method compares favorably with existing methods in simulations and an article sharing benchmark.

\end{abstract}


\section{Introduction}
\begin{figure}
\centering
\includegraphics[scale=.85]{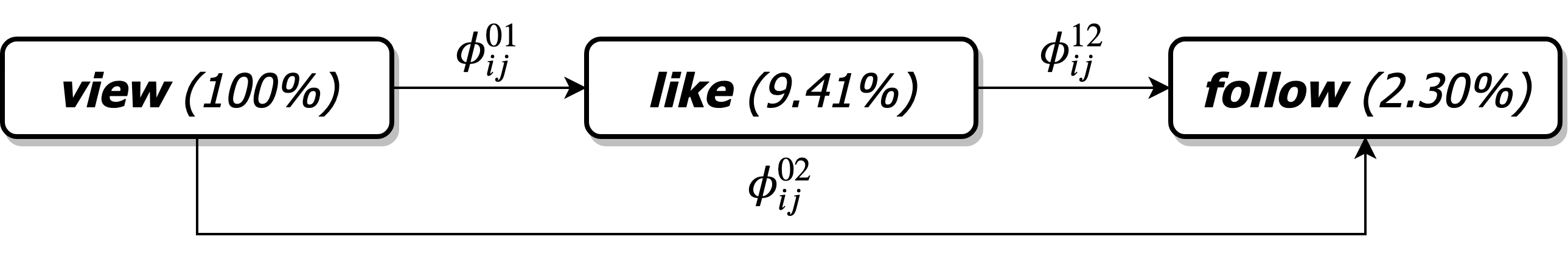}
\caption{Demonstration of multistage prediction for \textit{Deskdrop} dataset, a user may \textit{like} an article only if this user has \textit{viewed} it, and the user may \textit{follow} an article only if this user has \textit{liked} it. Three stagewise predictions based on different pairs of a given stage and a subsequent stage are considered: given \textit{view} to predict \textit{like}, given \textit{view} to predict \textit{follow}, and given \textit{like} to predict \textit{follow}. }
\label{fig:demo}
\end{figure}

A multistage recommender system on a monotonic chain of events predicts a user's preference of a large collection of items based on only a few user-item feedback at multiple stages, where a user's positive feedback of subsequent stages can only be observed given his/her positive feedback at present stages. As a result, a user may exhibit an increasing level of intention and preference given the feedback, ranging from the most implicit to the most explicit. It has been widely used for personalized prediction in e-commerce and social networks as well as individual drug responses over multiple phases in personalized medicine \citep{suphavilai2018predicting}. In such a situation, the objective of a multistage recommender system is to predict a user's subsequent behavior from his/her previous feedback with a high percentage of missing observations. For instance, as displayed in Figure \ref{fig:demo}, the \textit{Deskdrop} article sharing dataset\footnote{\url{https://www.kaggle.com/gspmoreira/articles-sharing-reading-from-cit-deskdrop}} consists of a sequence of actions ranging from \textit{view}, \textit{like}, to \textit{follow}, where only less than 0.2\% of values are observations with a total of $73,000$ users and more than 3,000 articles. A user \textit{likes} an article only if this user has \textit{viewed} it, and the user may \textit{follow} an article only when he/she has \textit{liked} it. Thus, three pairwise predictions are made based on three pairs of present and subsequent stages are performed, namely, the prediction of if the user will like an article given that he has \textit{viewed}, that if he/she will \textit{follow} given that he has \textit{viewed}, and that if he/she will \textit{follow} given that he has \textit{liked}.

One key characteristic of multistage prediction on a monotonic chain of events is a two-level monotonic property, as precisely defined in \eqref{eqn:two_side}. On the one hand, monotonicity occurs over subsequent stages given a present stage. For instance, based on the fact that a user has \textit{viewed} the article, this user won't \textit{follow} it if he/she doesn't \textit{like} it. On the other hand, monotonicity exhibits over present stages given a subsequent stage in a reversed order. That said, if the user has decided to \textit{follow} only he/she \textit{viewed} it, then the user will \textit{follow} an article when he/she has \textit{liked}.


\subsection{Literature review and our contributions}

Despite recent progress in recommender systems, multistage personalized prediction on a monotonic chain of events remains largely unexplored. A standard method is to predict a user's behaviors at a subsequent stage given a present stage individually, ignoring the monotonic property, as described in \eqref{eqn:indv}. In this sense, essentially all existing methods of recommender systems for a single-stage are applicable, including latent factor models \citep{koren2008factorization, dai2019smoothRS}, logistic matrix factorization \citep{johnson2014logistic}, tensor factorization\citep{Bi2018tensor}, and deep neural networks \citep{he2017neural}. The worth of note is that ordinal regression/classification may be adopted by treating subsequent stages as an ordered class given a present stage, as in \eqref{eqn:ord}. For example, in \cite{wan2018item, tran2012sequential}, latent factor models are developed based on an ordered logit model.
Yet, a two-level monotonic property introduced by a monotonic chain of events is not met, which considers only forward monotonicity of subsequent stages given the present one but not the backward monotonicity over present stages given the subsequent one. As to be seen, the two-level monotonicity yields not only a high accuracy of prediction but also prediction consistency \eqref{eqn:two_side} among different stage pairs. In practice, prediction consistency is indeed desirable for
decision-making. 

%


This article develops a multistage recommender system that models user-item-stage interactions and integrates the two-level monotonic property into personalized prediction on a monotonic chain of events. Three key contributions in this article can be summarized:

\begin{itemize}
\item The proposed method can produce a recommendation for any subsequent stage given observations at any present stage, which is highly demanded in real applications; see Figure \ref{fig:demo}. Yet, most conventional recommender systems focus on a fixed present stage.

\item A novel multistage loss function is proposed to treat multistage prediction for any pair of present and subsequent stages. This loss function admits user-item interactions observed at different stages and evaluates the prediction accuracy on all subsequent stages.

\item The two-level monotonic property is fully accounted for by our nonnegative additive latent factor model based on the Bayes rule in Lemma \ref{lem:bayes-rule}. As a result, it substantially reduces model parameters and most importantly ensures prediction consistency across different stages. By comparison, none of aforementioned methods can guarantee prediction consistency, c.f., Tables \ref{tab:sim} and \ref{tab:app}.

\item An algorithm is developed to implement the proposed method based on blockwise coordinate descent. Moreover, a learning theory is established to quantify the generalization error to demonstrate the benefits of modeling user-item-stage interactions based on the two-level monotonic property.

\end{itemize}

\section{Multistage classification on a monotonic chain}
 
Consider a recommender system in which triplets $(\Delta_{ij}, \bm{X}_{ij}, \bm{Y}_{ij} )$, with $\bm{Y}_{ij} = (Y^1_{ij}, \cdots, Y^T_{ij})$ are observed for user $i$ on item $j$ at stage $t$; $1 \leq i \leq n$, $1 \leq j \leq m$, $1 \leq t \leq T$. Here $n$ and $m$ are the number of users and items, respectively. $\Delta_{ij} = 1/0$ indicates if $(\bm X_{ij},\bm{Y}_{ij})$ is observed or missing and $\Omega_{t} = \{(i,j): \Delta_{ij} = 1, y^{t}_{ij}=1 \}$ is an index set of observed positive feedback at stage $t$, where $Y^t_{ij} = \pm 1$ with 1/-1 indicating positive/negative feedback at stage $t$, and $^\top$ denotes transpose. Moreover, $\bm{x}_{ij} =(\bm{u}_{i}^\top, \bm{v}_{j}^\top, \bm{s}_i^\top, \bm{o}^\top_j)^\top$ consists of both numerical and categorical features, where $\bm{u}_{i} \in [0,1]^{p_1}$ and $\bm{v}_j \in [0,1]^{p_2}$ are user-specific and item-specific numerical predictors, respectively, and normalization is performed for each feature to scale to $[0,1]$ otherwise, and moreover $\bm{s}_i = (s_{i1}, \cdots, s_{id_1})^\top$ with $s_{il} \in \{1, \cdots, n_l\}$ and $\bm{o}_j = (o_{j1}, \cdots, o_{jd_2})$ with $o_{jl} \in \{1, \cdots, m_l\}$ are $d_1$-dimensional and $d_2$-dimensional user-specific and item-specific categorical predictor vectors. For instance, in the \textit{Deskdrop} dataset\footnote{\url{https://www.kaggle.com/gspmoreira/articles-sharing-reading-from-cit-deskdrop}}, $\bm{v}_{j}$ is a numerical embedding for the content, $\bm{s}_i$ consists of ``person Id'', ``user agent'', ``user region'', ``user country'', $\bm{o}_j$ consists of ``content Id'', ``author Id'', ``language'', and $Y^1_{ij}$ and $Y^2_{ij}$ indicate if article $j$ is \textit{liked} and \textit{followed} by user $i$, respectively. One important characteristic of this kind of data is that subsequent events will not occur if one present event does not occur. For instance, if a user does not \textit{view} an item, then subsequent events of \textit{like} and \textit{follow} will not occur. 



On this ground, we define a monotonic behavior chain as follows:  
\begin{eqnarray}
\label{eqn:property}
 Y^t_{ij} = -1 \text{ if } Y^{t-1}_{ij} = -1; \quad 
1 \leq i \leq m, 1 \leq j \leq n, t=1,\cdots,T, 
\end{eqnarray}
which says that event $Y^{t-1}_{ij} = -1$ implies that any subsequent event $Y^{t}_{ij} = -1$ must occur, which encodes a certain causal relation as defined by the local Markov property \citep{edwards2012introduction} in a directed acyclic graphical model.

\subsection{Two-level monotonicity}

Multistage classification learns to predict the outcome of $Y_{ij}^t$ at a subsequent stage $t$ given observations at a present stage $Y_{ij}^{t'}$ and $\bm{x}_{ij}$; $0 \leq t'<t \leq T$. With the convention, we set $Y^0_{ij} = 1$. To predict, we introduce a decision function $\phi_{ij}^{t' t} = \phi_{ij}^{t' t} (\bm{x}_{ij}, y^{t'}_{ij})$ for classification, which depends on observations $\bm{x}_{ij}$ and $y^{t'}_{ij}$. For user $i$ on item $j$ across $T$ stages, decision functions can be expressed as $\bm{\phi}_{ij} = (\phi_{ij}^{t't})_{0\leq t' < t \leq T}$. For all users over items across $T$ stages, decision functions are $\bm{\phi}=(\bm{\phi}_{ij})_{1 \leq i \leq n; 1 \leq j \leq m}$.

To evaluate the overall performance of $\bm{\phi}$, we define the multistage misclassification error
\begin{eqnarray}
\label{eqn:loss}
e(\bm \phi) = (nm)^{-1}\sum_{i=1}^n \sum_{j=1}^m e_{ij}(\bm{\phi}_{ij}), \quad 
e_{ij}(\bm{\phi}_{ij})  =  \sum_{0 \leq t' < t \leq T} w_{t' t} \mathbb{E}_{ij}\big( \Delta_{ij} 
\mathbb{I} \big(Y_{ij}^t \bm{\phi}_{ij}^{t't}(\bm{X}_{ij}, Y^{t'}_{ij}) \leq 0 \big) \big),
\end{eqnarray}
where $e_{ij}(\bm{\phi}_{ij})$ is the pairwise misclassification error for user $i$ on item $j$ across $T$ stages, and $\mathbb{P}_{ij} (\cdot) = \mathbb{P}_{ij}(\cdot \mid \bm{X}_{ij} = \bm{x}_{ij})$ and $\mathbb{E}_{ij} (\cdot) = \mathbb{E} (\cdot \mid \bm{X}_{ij} = \bm{x}_{ij})$ denote the conditional probability and expectation given $\bm{X}_{ij} = \bm{x}_{ij}$. In \eqref{eqn:loss}, $w_{t' t} \geq 0$ is a pre-specified weight for predicting the outcome at subsequent stage $t$ based on present stage $t'$; $0 \leq t'<t \leq T$, reflecting the relative importance with respect to prediction at different stages in the overall evaluation. In particular, \eqref{eqn:loss} reduces to the next-stage prediction and last-stage prediction when $w_{t' t} = I(t - t' = 1)$ and $w_{t't} = I(t = T)$, respectively. Moreover, if missing occurs completely at random, or missing pattern $\Delta_{ij}$ is independent of $\bm{Y}_{ij}$, then \eqref{eqn:loss} is proportional to the standard misclassification error.  


Lemma \ref{lem:bayes-rule} gives the multistage Bayes decision function $\bar{\bm{\phi}}_{ij}=\argmin_{\bm{\phi}_{ij}} e_{ij}(\bm{\phi}_{ij})$ for user $i$ on item $j$ in \eqref{eqn:loss} subject to a constraint that predicted $Y_{ij}^t$ by the Bayes rule satisfies the monotonic behavior chain property \eqref{eqn:property}.

\begin{lemma}[Multistage Bayes-rule] \label{lem:bayes-rule} The optimal multistage pairwise decision function ${\bar{\bm \phi}}_{ij}$ minimizing \eqref{eqn:loss} can be written as: for $1\leq i \leq n$, $1 \leq j \leq m$, $0 \leq t' < t \leq T$,  
\begin{eqnarray*}
\bar{\phi}_{ij}^{t't}(\bm{x}_{ij}, y^{t'}_{ij}) = \left\{
		\begin{array}{ll}
      -1, \quad \text{if } y^{t'}_{ij} = -1, \\
      \sign\big(\bar{f}_{ij}^{t't}(\bm{x}_{ij})\big) = \sign\big( \mathbb{P}_{ij}\big( Y^t_{ij} = 1 | Y^{t'}_{ij} = 1, \Delta_{ij} = 1 \big) - 1/2 \big), \quad \text{if }   y^{t'}_{ij} = 1;
    \end{array}
  \right.
\end{eqnarray*}
The multistage Bayes rule for predicting $Y^t_{ij}$ by $\bm{x}_{ij}$ and $Y^{t'}_{ij}$ is $\sign \big(\bar{\phi}_{ij}^{t' t}(\bm x_{ij},Y^{t'}_{ij}) \big)$. Here $\bar{f}_{ij}^{t't}(\bm{x}_{ij})$ satisfies the two-level monotonic property \eqref{eqn:two_side} iff $Y^t_{ij}$ follows \eqref{eqn:property}, that is, for $0 \leq t' < t \leq T$, 
\begin{align}
\label{eqn:two_side}
& \text{Forward:} & \sign(\bar{f}_{ij}^{t'(t+1)}(\bm{x}_{ij})) = -1, \text{ if } \sign(\bar{f}_{ij}^{t't}(\bm{x}_{ij})) = -1,  \nonumber \\
& \text{Backward:} & \sign(\bar{f}_{ij}^{t't}(\bm{x}_{ij})) = -1,  \text{ if } \sign(\bar{f}_{ij}^{(t'+1)t}(\bm{x}_{ij})) = -1.
\end{align}
Moreover, there exists $\bar{h}^r_{ij}(\bm x_{ij}) \geq 0$ such that $\bar{f}^{t't}_{ij}(\bm{x}_{ij})$ can be written in an additive form: 
\begin{equation}
\bar{f}^{t't}_{ij}(\bm{x}_{ij}) = \bar{h}^0_{ij}(\bm{x}_{ij}) - \sum_{r = t'+1}^t \bar{h}^r_{ij}(\bm{x}_{ij}), \text{ with } \bar{h}^{r}_{ij} \geq 0; \ 0 \leq r \leq T.
\label{eqn:additive}
\end{equation}
\end{lemma}

The two-level monotonic property \eqref{eqn:two_side} says that $\sign(\bar{f}_{ij}^{t't}(\bm{x}_{ij}))$ is decreasing in $t$ for any fixed $t'$ (forward monotonicity) whereas is increasing in $t'$ for any fixed $t$ (backward monotonicity). Note that \eqref{eqn:additive} guarantees \eqref{eqn:two_side}. 

In view of  Lemma \ref{lem:bayes-rule}, we introduce our multistage pairwise decision functions to mimic the additive multistage Bayes rule: for some $h_{ij}^r(\bm{x}_{ij}) \geq 0$; $0 \leq r \leq T$.
\begin{eqnarray}
\label{eqn:decision}
   & \phi_{ij}^{t't}(\bm{x}_{ij}, y_{ij}^{t'}) =  \left\{
           		\begin{array}{ll}
                  -1, \quad \text{if } y^{t'}_{ij} = -1, \\
                  f_{ij}^{t't}(\bm{x}_{ij}) = h_{ij}^{0}(\bm{x}_{ij}) - \sum_{r=t'+1}^t h_{ij}^{r}(\bm{x}_{ij}), \quad \text{if} \ y^{t'}_{ij} = 1.
                \end{array}
              \right.
\end{eqnarray}


\subsection{Two-level monotonic multistage classification}

Based on the representation of decision function in \eqref{eqn:decision}, we rewrite $e_{ij}(\bm{\phi}_{ij})$ in \eqref{eqn:loss} as follows:
\begin{align}
\label{eqn:reduce_loss}
   e_{ij}(\bff_{ij})  
  & = \sum_{0 \leq t' < t \leq T} w_{t' t} \mathbb{E}_{ij}\big( \Delta_{ij} \mathbb{I} \big(Y^t_{ij} f^{t't}_{ij}(\bm{X}_{ij}) \leq 0  \big) \mathbb{I}(Y^{t'}_{ij} = 1) \big).
\end{align}
Note that the indicator function $\mathbb{I}(\cdot)$ in \eqref{eqn:reduce_loss} is difficult to treat in optimization. Therefore, we replace it by a surrogate loss $V(u)$ for large-margin classification, in which $V$ is a function of the corresponding functional margin $Y^t_{ij} f^{t't}_{ij}(\bm{x}_{ij})$. This includes, but is not limited to, the hinge loss $V(u) = (1 - u)_+$ \citep{cortes1995support}, the import vector machine $V(z)=\log(\exp(-z)/(1-\exp(-z)))$ \citep{zhu2002kernel}, and the $\psi$-loss $V(u) = \min(1, (1-u)_+)$ \citep{shen2003psi}.
On this ground, we propose a multistage large-margin loss function:
\begin{align}
\label{eqn:psi_loss}
L(\bff_{ij}, \bm Z_{ij}) & = \sum_{0 \leq t'< t \leq T} w_{t' t} \Delta_{ij} 
V \big(Y^t_{ij} f^{t't}_{ij}(\bm{x}_{ij}) \big) \mathbb{I}(Y^{t'}_{ij} = 1), 
\end{align}
where $\bm Z_{ij} = (\Delta_{ij}, \bm Y_{ij})$ and
$\bff_{ij}=(f_{ij}^{t' t})_{0\leq t^{'} < t \leq T}$. 

Lemma \ref{lem:FC} says that a minimizer of the cost $l(\bff) = (nm)^{-1}\sum_{ij} 
\mathbb{E}_{ij} L(\bff_{ij}, \bm{Z}_{ij})$ with respect to $\bff$ satisfies the Bayes rule in Lemma \ref{lem:bayes-rule}.

\begin{lemma}[Multistage Fisher consistency]
\label{lem:FC} The minimizer of $l(\bff)$ is Fisher-consistent in that 
it satisfies the Bayes rule in Lemma 1 if the surrogate loss $V(\cdot)$ is Fisher consistency in binary classification.
\end{lemma}

Next we parametrize our decision functions based on an additive latent factor model with to incorporate the collaborative information across users, items and stages. In particular, we define $h^r_{ij}(\bm{x}_{ij}) = \big( a( \bm{u}_i, \bm{s}_i ) \circ b(\bm{v}_j, \bm{o}_j )\big)^\top \bm{q}_r$ in \eqref{eqn:additive}, and set $\bm{q}_0 = \bm{1}$ to avoid the over-parametrization. Moreover, $a( \bm{u}_i, \bm{s}_i )$ and $b(\bm{v}_j, \bm{o}_j )$ are proposed to link our decision function linearly to numerical predictors $\bm{u}_{i}$ and $\bm{v}_j$, as well as  additive latent factors structured by categorical predictors $\bm{s}_i$ and $\bm{o}_j$. Then, the proposed prediction function can be written as
\begin{align}
\label{eqn:decision_function}
& f_{ij}^{t' t}(\bm{x}_{ij}) = a^\top( \bm{u}_i, \bm{s}_i ) b(\bm{v}_j, \bm{o}_j )  - \sum_{r = t'+1}^t \big( a( \bm{u}_i, \bm{s}_i ) \circ b(\bm{v}_j, \bm{o}_j )\big)^\top \bm{q}_r,  \\
& \ a(\bm{u}_i, \bm{s}_i) = \bm{A} \bm{u}_i + \sum_{l=1}^{d_1}\bm{a}_{ls_{il}}, \ b(\bm{v}_j, \bm{o}_j) = \bm{B} \bm{v}_j + \sum_{l=1}^{d_2} \bm{b}_{lo_{jl}}, \nonumber 
\end{align}
subject to $\bm{A} \geq \bm{0}$, $\bm{B} \geq \bm{0}$, $\bm{q}_r \geq \bm{0}$; $r = 1, \cdots, T$, $\bm{a}_{lh} \geq \bm 0$; $l = 1, \cdots, d_1$; $h = 1, \cdots, n_l$; $\bm{b}_{lh} \geq \bm 0$; $l = 1, \cdots, d_2$; $h = 1, \cdots, m_l$, where $\circ$ is the Hadamard product, $\bm{A} \in \mathbb{R}^{K \times p_1}$ and $\bm{B} \in \mathbb{R}^{K \times p_2}$ are two matrices, which transform the user-specific and item-specific numerical features to $K$-dimensional latent vectors, and $\bm{a}_{ls_{il}}$ and $\bm{b}_{lo_{jl}}$are two $K$-dimensional latent factor for $s_{il}$ and $o_{jl}$.



Representation \eqref{eqn:decision_function} is highly interpretable, it leverages all feature-interaction between users and items based on an additive model. For example, in Deskdrop dataset, the interaction effect of ``userId''-``articleId'', ``userId''-``authorId'', ``userAgent''-``articleId'', ``userAgent''-``authorId'', are all captured in \eqref{eqn:decision_function}.
Furthermore,  nonnegative constraints for $\bm{A} \geq \bm{0}$, $\bm{B}\geq \bm{0}$, $\bm{a}_{lh} \geq \bm 0$ and $\bm{b}_{lh} \geq \bm 0$ are enforced to ensure the two-level monotonic property \eqref{eqn:two_side}.

\begin{figure}
\centering
\includegraphics[scale=.55]{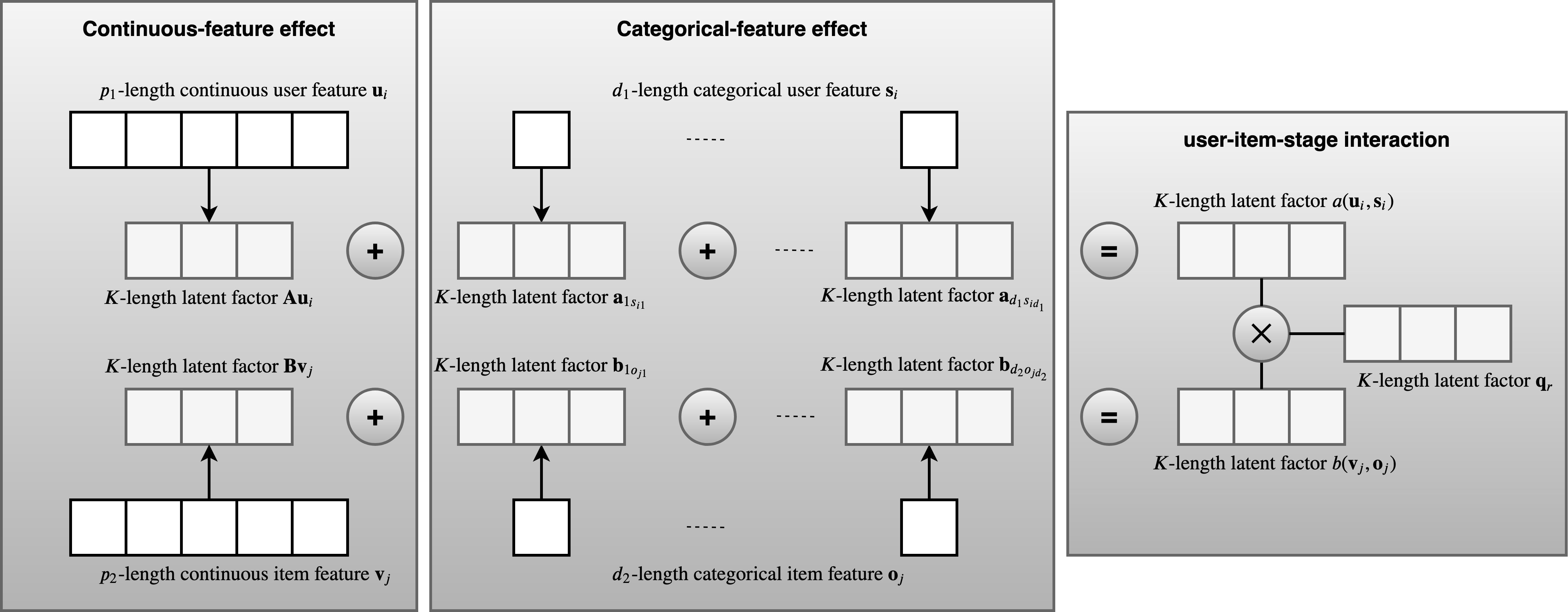}
\caption{Architecture of proposed nonnegative additive latent factor models for continuous and categorical user-item specific features as well as the user-item-stage effect.} \label{fig:classifier}
\end{figure}

Given observations $(\bm{x}_{ij},y^t_{ij})_{(i,j)\in \Omega_0, 1 \leq t \leq T}$ and positive index set $\Omega_{t'} = \{(i,j): \Delta_{ij} = 1, y^{t'}_{ij} = 1 \}$ at stage $t'$, we propose a regularized multistage large-margin loss
\begin{align}
\label{model}
\min_{\mathbf{f} \in \mathcal{F}} \tilde{l}_{nm}(\mathbf{f}); & \quad \tilde{l}_{nm}(\mathbf{f})=(nm)^{-1} \sum_{0 \leq t'< t \leq T} \sum_{(i,j) \in \Omega_{t'}} 
 w_{t' t} V \big(y^t_{ij} f^{t't}_{ij}(\bm{x}_{ij}) \big)
 + J_{\lambda}(\mathbf{f}),
\end{align}
subject to $\bm{A} \geq \bm{0}$, $\bm{B} \geq \bm{0}$, $\bm{a}_{lh} \geq 0$; $1 \leq l \leq d_1$; $1 \leq h \leq n_l$, $\bm{b}_{lh} \geq 0$; $1 \leq l \leq d_2$,$1 \leq h \leq m_l$, where $J_{\lambda}(\mathbf{f}) = \lambda_1 \big(\|\bm{A}\|_F^2 + \|\bm{B}\|_F^2 \big) + \lambda_2 \big(\| \bm{a} \|_2^2 + \| \bm{b} \|_2^2 \big) + \lambda_3 \| \bm{q} \|_2^2 $, $\bm{a} = (\bm{a}^\top_{1}, \cdots, \bm{a}^\top_{d_1})^\top$, $\bm{b} = (\bm{b}^\top_{1}, \cdots, \bm{b}^\top_{d_2})^\top$, $\bm{q} = (\bm{q}^\top_1, \cdots, \bm{q}^\top_T)^\top$, $\bm{a}_{l} = (\bm{a}_{l1}^\top, \cdots, \bm{a}_{ln_l}^\top)^\top$, $\bm{b}_{l} = (\bm{b}_{l1}^\top, \cdots, \bm{b}_{lm_l}^\top)^\top$, $\lambda_1, \lambda_2, \lambda_3 \geq 0$ are tuning parameters controlling the trade-off between learning and regularization, and $\mathcal{F}$ is a space of candidate decision functions in \eqref{eqn:decision_function}, which can be written as
\begin{align}
& \mathcal{F} = \Big\{\mathbf{f} = (\mathbf{f}^{t't}_{ij}): \mathbf{f}^{t't}_{ij}(\bm{x}_{ij}) = a^\top( \bm{u}_i, \bm{s}_i ) b(\bm{v}_j, \bm{o}_j )  - \sum_{r = t'+1}^t \big( a( \bm{u}_i, \bm{s}_i ) \circ b(\bm{v}_j, \bm{o}_j )\big)^\top \bm{q}_r, \nonumber \\
& \bm{a}_{lh} \in \mathbb{R}^K_+, \quad  l = 1, \cdots, d_1, h = 1, \cdots, n_l; \quad 
 \bm{b}_{lh} \in \mathbb{R}^K_+, \quad  l = 1, \cdots, d_2, h = 1, \cdots, m_l, \nonumber \\
& \bm{q}_t \in \mathbb{R}_+^K, \quad t = 1, \cdots, T; \quad \bm{A} \in \mathbb{R}_+^{K \times p_1}, \bm{B} \in \mathbb{R}_+^{K \times p_2} \Big\}, 
\end{align}
where $a(\cdot, \cdot)$ and $b(\cdot, \cdot)$ are defined in \eqref{eqn:decision_function}.

Minimization \eqref{model} in $(\bm A,\bm B, \bm q, \bm b, \bm q)$ yields an estimated $(\hat{\bm{A}}, \hat{\bm{B}}, \hat{\bm{a}}, \hat{\bm{b}}, \hat{\bm{q}})$ thus $\widehat{\bff}_{ij}^{t' t}$ by \eqref{eqn:decision_function} and $\widehat{\bm{\phi}}_{ij}^{t' t}$ by \eqref{eqn:decision}. Finally, the outcome of $Y_{ij}^{t}$ is predicted by $\sign(\widehat{\bm{\phi}}_{ij}^{t' t}(\bm x_{ij},Y_{ij}^{t'}))$; $1 \leq i \leq n$, $1 \leq j \leq m$, $0 \leq t' < t \leq T$. 

\subsection{Connection with existing frameworks}
   
This section compares the proposed method with a standard method treating each stage individually and an ordinal method treating the stage as an ordinal class. 

\textbf{Standard.} A standard method treats $(t', t)$-pairwise classification separately and then combine the prediction results for $0 \leq t' < t \leq T$. In fact, it estimates $\mathbf{f}^{t't} = (f_{ij}^{t't})_{1 \leq i \leq n; 1 \leq j \leq m}$ by solving: 
\begin{align}
\label{eqn:indv}
\min_{\mathbf{f}^{t't} \in \mathcal{F}^{t't}} \ (nm)^{-1} \sum_{(i,j) \in \Omega_{t'}} V \big(y^t_{ij} f^{t't}_{ij}(\bm{x}_{ij}) \big) + J_{\lambda}(\mathbf{f}^{t't}); \quad 0 \leq t' < t \leq T,
\end{align}
where $\mathcal{F}^{t't}$ is a parameter space of candidate pairwise decision functions. A standard method includes latent factor models \citep{koren2008factorization}, gradient boosting \citep{cheng2014gradient}, and deep neural networks \citep{he2017neural}, and $J_{\lambda}(\mathbf{f}^{t't})$ regularizes $\mathbf{f}^{t't}$ with a nonnegative tuning parameter $\lambda \geq 0$.

\textbf{Ordinal.} An ordinal method treats each stage $t'$ separately, which estimates $\mathbf{f}^{t'} = (\mathbf{f}^{t't})_{t' < t \leq T}$ by solving: 
\begin{align}
\label{eqn:ord}
& \min_{\mathbf{f}^{t'} \in \mathcal{F}^{t'}} \ (nm)^{-1}  \sum_{(i,j) \in \Omega_{t'}} \sum_{t = t'+1}^T w_{t't} V \big(y^t_{ij} f^{t't}_{ij}(\bm{x}_{ij}) \big) + J_{\lambda}(\mathbf{f}^{t'}), \quad \text{for } 0 \leq t' < T, 
\end{align}
subject to $f^{t'(t'+1)}_{ij} \geq f^{t'(t'+2)}_{ij} \geq \cdots \geq f^{t'T}_{ij}$, where $\mathcal{F}^{t'}$ is a parameter space of candidate stagewise decision functions and $J_{\lambda}(\mathbf{f}^{t'})$ regularizes $\mathbf{f}^{t'}$. For example, \cite{crammer2002pranking} formulates a parallel decision function $f_{ij}^{t't}(\bm{x}_{ij}) = \bm{\beta}^\top_{t'} \bm{x}_{ij} + \beta_{t't}$ and imposes the constraint $\beta_{t'(t'+1)} \geq \beta_{t'(t'+2)} \cdots \geq \beta_{t'T}$. Moreover, since $\bm{x}_{ij}$ is pre-scaled in $[0,1]$, we can set $\mathbf{f}^{t't}_{ij} = (\bm{\beta}_{t'0} - \sum_{t=t'+1}^{T}\bm{\beta}_{t't})^\top \bm{x}_{ij}$, where the forward monotonicity is ensured by positive constraints $\bm{\beta}_{t't} \geq \bm{0}$, $t = t'+1, \cdots, T$.  Note that when $V(y^t_{ij} f^{t't}_{ij})$ is replaced as a negative log-likelihood function, then \eqref{eqn:ord} is a formulation for ordinal regression \cite{mccullagh2019generalized, bhaskar2016probabilistic}.

Ordinal classification incorporates the forward monotonicity but ignores backward monotonicity. Whereas such monotonicity helps to reduce the size of the parameter space, by comparison, the proposed method can further reduce parameters utilizing the backward monotonicity. In contrast, a standard method does not leverage any level of monotonicity. Most critically, both standard and ordinal methods fail to yield prediction consistency.

\section{Large-scale computation}


This section develops a computational scheme to solve nonconvex minimization \eqref{model}. For illustration, consider the hinge loss $V(u) = (1-u)_+$. The scheme minimizes a nonconvex cost function \eqref{model} by solving a sequence of relaxed convex subproblems via a block successive minimization (BSM). 
The scheme uses blockwise descent to alternate the following convex subproblems. For each subproblem, we decompose \eqref{model} into an equivalent form of many small optimizations for parallelization and for alleviating the memory requirement. Let $\bm{q}_{t't} = \bm{1}_K - \sum_{r = t'+1}^t \bm{q}_r$, we solve \eqref{model} as follows.



\textbf{User-effect block $\bm{A}$.} This convex optimization solves for $\bm{A}$:
\begin{align}
\label{opt:user_A}
\min_{\text{vec}(\bm{A}) \geq 0} & \ (nm)^{-1} \sum_{0 \leq t' < t \leq T} \sum_{(i,j) \in \Omega_{t'}} w_{t't} \xi_{ij}^{t't} + \lambda_1 \sum_{h=1}^{n_l} \| \text{vec}(\bm{A}) \|_2^2, \nonumber \\
& \xi_{ij}^{t't} \geq 1 - y^t_{ij} \Big( \text{vec}(\bm{A})^\top \big( \bm{u}_i \otimes \big( b(\bm{v}_j, \bm{o}_j) \circ \bm{q}_{t't} \big) \big) + \big( \sum^{d_1}_{l=1} \bm{a}_{ls_{il}} \big)^\top \big(  b(\bm{v}_j, \bm{o}_j) \circ \bm{q}_{t't} \big) \Big), \nonumber \\
& \xi_{ij}^{t't} \geq 0; (i,j) \in \Omega_{t'}, \ 0 \leq t' < t \leq T,
\end{align}
where $\otimes$ is the Kronecker product and $\text{vec}(\bm{A})$ is the column vectorization of 
matrix $\bm{A}$.

\textbf{User-effect block $\bm{a}$.} This convex optimization solves for $(\bm{a}_{1}, \cdots,  \bm{a}_{d_1})$:
\begin{align}
\label{opt:user_all}
\min_{\bm{a}_l \geq \bm{0}} & \ (nm)^{-1} \sum_{h=1}^{n_l}  \sum_{0 \leq t' < t \leq T} \sum_{\{i: \bm{s}_{il} = h \}} \sum_{j \in \{j: (i,j) \in \Omega_{t'}\}}  w_{t't} \xi_{ij}^{t't} + \lambda_2 \sum_{h=1}^{n_l} \| \bm{a}_{lh} \|_2^2, \nonumber \\
& \xi_{ij}^{t't} \geq 1 - y^t_{ij} \Big( \bm{a}^\top_{ls_{il}} \big( b(\bm{v}_j, \bm{o}_j) \circ \bm{q}_{t't} \big) + \big( \bm{A}\bm{u}_i + \sum_{l' \neq l} \bm{a}_{l's_{il'}} \big)^\top \big( b(\bm{v}_j, \bm{o}_j) \circ \bm{q}_{t't} \big) \Big), \nonumber \\
& \xi_{ij}^{t't} \geq 0; (i,j) \in \Omega_{t'}; \ 0 \leq t' < t \leq T.
\end{align}
Note that \eqref{opt:user_all} can be separately solved for each $\bm{a}_{lh}$ in a parallel fashion,
\begin{align}
\label{opt:user_indv}
 \min_{\bm{a}_{lh} \geq \bm{0}} & \ (nm)^{-1}   \sum_{0 \leq t'<t \leq T} \sum_{\{i: \bm{s}_{il} = h \}} \sum_{j \in \{j:(i,j) \in \Omega_{t'} \} }  w_{t't} \xi_{ij}^{t't} + \lambda_2 \| \bm{a}_{lh} \|_2^2, \nonumber \\ 
& \xi_{ij}^{t't} \geq 1 - y^t_{ij} \Big( \bm{a}^\top_{lh} \big( b(\bm{v}_j,\bm{o}_j) \circ \bm{q}_{t't} \big) + \big( \bm{A}\bm{u}_i + \sum_{l' \neq l} \bm{a}_{l's_{il'}} \big)^\top \big( b(\bm{v}_j, \bm{o}_j) \circ \bm{q}_{t't} \big) \Big), \nonumber \\
& \xi_{ij}^{t't} \geq 0; \ (i,j) \in \Omega_{t'}; \ 0 \leq t' < t \leq T.
\end{align} 

\textbf{Item-effect block $\bm{B}$.} This convex optimization solves for $\bm{B}$: 
\begin{align}
\label{opt:item_B}
\min_{\text{vec}(\bm{B}) \geq 0} & \ (nm)^{-1}  \sum_{0 \leq t' < t \leq T} \sum_{(i,j) \in \Omega_{t'}} w_{t't} \xi_{ij}^{t't} + \lambda_1 \sum_{h=1}^{n_l} \| \text{vec}(\bm{B}) \|_2^2, \nonumber \\
& \xi_{ij}^{t't} \geq 1 - y^t_{ij} \Big( \text{vec}(\bm{B})^\top \big( \bm{v}_j \otimes \big( a(\bm{u}_i, \bm{s}_j) \circ \bm{q}_{t't} \big) \big) + \big( \sum^{d_1}_{l=1} \bm{b}_{ls_{il}} \big)^\top \big( a(\bm{u}_i, \bm{s}_i) \circ \bm{q}_{t't} \big) \Big), \nonumber \\
& \xi_{ij}^{t't} \geq 0; (i,j) \in \Omega_{t'}, \ 0 \leq t' < t \leq T.
\end{align}

\textbf{Item-effect block $\bm{b}$.} For $l = 1, \cdots, d_2$,  $(\bm{b}_{lh})_{h=1}^{m_l}$ is solved in a parallel fashion,
\begin{align}
\label{opt:item_b}
 \min_{\bm{b}_{lh} \geq \bm{0}} & \ (nm)^{-1}   \sum_{0 \leq t'<t \leq T} \sum_{\{j: \bm{o}_{jl} = h \}} \sum_{i \in \{i: (i,j) \in \Omega_{t'} \}}   w_{t't} \xi_{ij}^{t't} + \lambda_2 \| \bm{b}_{lh} \|_2^2, \nonumber \\ 
& \xi_{ij}^{t't} \geq 1 - y^t_{ij} \Big( \bm{b}^\top_{lh} \big( a(\bm{u}_i,\bm{s}_i) \circ \bm{q}_{t't} \big) + \big( \bm{B}\bm{v}_j + \sum_{l' \neq l} \bm{b}_{l'o_{jl'}} \big)^\top \big( a(\bm{u}_i, \bm{s}_i) \circ \bm{q}_{t't} \big) \Big), \nonumber \\
& \xi_{ij}^{t't} \geq 0; j \in \{j: \bm{o}_{jl} = h\}, \ (i,j) \in \Omega_{t'}; \ 0 \leq t' < t \leq T.
\end{align} 

\textbf{Stage-effect block $\bm{q}$.} This convex optimization solves for $\bm{q}_r$, for $r = 1, \cdots, T$, given the present values for all other variables:
\begin{align}
\label{opt:stage}
\min_{\bm{q}_r \geq \bm{0}} & \ (nm)^{-1}  \sum_{0 \leq t'< r \leq t \leq T} \sum_{(i,j) \in \Omega_{t'}}  w_{t't} \xi_{ij}^{t't} + \lambda_3 \| \bm{q}_{r} \|_2^2, \nonumber \\ 
& \xi_{ij}^{t't} \geq 1 - y^t_{ij} \Big( - \bm{q}_r^\top \big( a(\bm{u}_i, \bm{s}_i) \circ b(\bm{v}_j, \bm{o}_j) \big) + (\bm{1} - \sum_{r'=t'+1; r' \neq r}^t \bm{q}_r)^\top \big( a(\bm{u}_i, \bm{s}_i) \circ b(\bm{v}_j, \bm{o}_j) \big) \Big), \nonumber \\
& \xi_{ij}^{t't} \geq 0; (i,j) \in \Omega_{t'}; \ 0 \leq t' < t \leq T.
\end{align} 

As a technical note, \eqref{opt:user_A}-\eqref{opt:stage} are standard SVMs with fixed intercept and positive constrains for model parameters, which can be solved via one efficient implementation in our python package \texttt{varsvm}\footnote{\url{https://pypi.org/project/varsvm/}} based on the coordinate decent algorithm \citep{wright2015coordinate}. 

The aforementioned scheme is summarized in Algorithm 1. 

\noindent \textbf{Algorithm 1 (Parallelized version: hinge loss)} 

\noindent \textbf{Step 1} (Initialization). Initialize the values of ($\bm{A}$, $\bm{B}$, $\bm{a}$, $\bm{b}$, $\bm{q}$) and specify the tolerance error.



\noindent \textbf{Step 2} (Update $\bm{A}$ and $\bm a$). Update $\bm{A}$ by solving 
\eqref{opt:user_A} given present values of the other variables.
For $l = 1, \cdots, d_1$, update $\bm{a}_{lh}$; $h=1, \cdots, n_l$ in a parallel fashion by solving \eqref{opt:user_indv} given present values of the other variables.


\noindent \textbf{Step 3} (Update $\bm{B}$ and $\bm b$). Update $\bm{B}$ by solving \eqref{opt:item_B} given present values of the other variables.
For $l = 1, \cdots, d_2$, update $\bm{b}_{lh}$; $h=1, \cdots, m_l$ in a parallel fashion by solving \eqref{opt:item_b} given present values of the other variables. 

\noindent \textbf{Step 4} (Update $\bm{q}$). For $r = 1, \cdots, T$, update $\bm{q}_{r}$ by solving \eqref{opt:stage} given present values of the other variables. 

\noindent \textbf{Step 5.} Iterate \textbf{Steps 2-4} until the decrement of the cost function in \eqref{model} is less than the tolerance error.

Algorithm 1 returns an estimate $(\widehat{\bm{A}}, \widehat{\bm{B}}, \widehat{\bm{a}}, \widehat{\bm{b}}, \widehat{\bm{q}})$ at termination, which in turn yields an estimated decision functions $\widehat{f}^{t't}_{ij}(\bm{x}_{ij}) = \widehat{a}^\top( \bm{u}_i, \bm{s}_i ) \widehat{b}(\bm{v}_j, \bm{o}_j )  - \sum_{r = t'+1}^t \big( \widehat{a}( \bm{u}_i, \bm{s}_i ) \circ \widehat{b}(\bm{v}_j, \bm{o}_j )\big)^\top \widehat{\bm{q}}_r$, and $\widehat{\phi}^{t't}_{ij} = -1$ if $Y^{t'}_{ij} = -1$, $\widehat{f}^{t't}_{ij}$ otherwise. Finally, the prediction at the $t$-th stage given the $t'$-th stage is $\sign \big(\widehat{\phi}^{t't}_{ij}(y^{t'}_{ij}, \bm{x}_{ij}) \big)$ for new user-item covariates $\bm{x}_{ij}$.

To implement Algorithm 1, we develop a Python library to solve an SVM-type of optimization, including weighted SVMs, drifted SVMs, and non-negative SVMs, based on coordinate descent of the dual problem \citep{wright2015coordinate}. This implementation has been available in \texttt{varsvm} library in Python on the GitHub repository \texttt{varsvm}\footnote{\url{https://github.com/statmlben/variant-svm}}, and the source code is released with the MIT license using GitHub and the release is available via PyPI. 
The development undergoes an integration of Ubuntu Linux and Mac OS X, in Python 3.

\begin{lemma}[Convergence of Algorithm 1]
\label{lem:algo}
A solution $(\widehat{\bm{A}}, \widehat{\bm{a}}, \widehat{\bm{B}}, \widehat{\bm{b}}, \widehat{\bm{q}})$ of Algorithm 1 is a stationary point of $\tilde{l}_{nm}$ 
in \eqref{model} in that
\begin{align}
	\tilde{\bm{A}} & = \argmin_{\bm{A}} \tilde{l}_{nm}(\bm{A}, \tilde{\bm{a}}, \tilde{\bm{B}}, \tilde{\bm{b}}, \tilde{\bm{q}}), \quad \tilde{\bm{B}} = \argmin_{\bm{B}} \tilde{l}_{nm}(\tilde{\bm{A}}, \tilde{\bm{a}}, \bm{B}, \tilde{\bm{b}}, \tilde{\bm{q}}), \nonumber \\
	\tilde{\bm{a}}_{l} & = \argmin_{\bm{a}_{l}} \tilde{l}_{nm}(\tilde{\bm{A}}, \tilde{\bm{a}}_{1}, \cdots, \bm{a}_{l}, \cdots, \tilde{\bm{a}}_{d_1}, \tilde{\bm{B}}, \tilde{\bm{b}}, \tilde{\bm{q}}); \ 1 \leq l \leq d_1; \nonumber \\
	\tilde{\bm{b}}_{l} & = \argmin_{\bm{b}_{l}} \tilde{l}_{nm}(\tilde{\bm{A}}, \tilde{\bm{a}}, \tilde{\bm{B}}, \tilde{\bm{b}}_{1}, \cdots, \bm{b}_{l}, \cdots, \tilde{\bm{b}}_{d_2}, \tilde{\bm{q}}); \  1 \leq l \leq d_2; \nonumber \\
	\tilde{\bm{q}}_t & = \argmin_{\bm{q}_t} \tilde{l}_{nm}(\tilde{\bm{A}}, \tilde{\bm{a}}, \tilde{\bm{B}}, \tilde{\bm{b}}, \tilde{\bm{q}}_1, \cdots, \bm{q}_t, \cdots, \tilde{\bm{q}}_T); \ 1 \leq t \leq T. \nonumber
\end{align}
\end{lemma}


The successive update in Algorithm 1 suffices to ensure its convergence because the cost function in \eqref{model} is strictly blockwise convex, that is, \eqref{opt:user_A}, \eqref{opt:user_indv}, \eqref{opt:item_B}, \eqref{opt:item_b}, and \eqref{opt:stage} are strictly convex in terms of its parameters within each block, yielding a unique minimizer at each step. This aspect differs from the maximum block improvement \cite{chen2012maximum} that guarantees the convergence of blockwise coordinate descent for a general objective function. One benefit of successive updating is that it significantly reduces computational complexity. Note that the solution of Algorithm 1 can be a global minimizer when certain additional assumptions are made \cite{haeffele2015global}.

\section{Theory}
This section investigates the generalization aspect of the proposed multistage recommender $\hat{\mathbf{f}}$ in terms of the accuracy of classification, as measured by the classification regret, defined as $e(\hat{\mathbf{f}}) - e(\bar{\mathbf{f}})$, where $e(\cdot)$ is the generalization error defined in \eqref{eqn:loss}.  


Assume that $(\Delta_{ij}, Y_{ij})$ given $\bm{X}_{ij}$; $1 \leq i \leq n$; $1 \leq j \leq m$, are conditionally independent, although $(\Delta_{ij}, \bm{X}_{ij}, Y_{ij})$'s may not be independent as multiple items can be purchased for a same user or multiple users may purchase a same item. Let $\bar{\mathbf{f}} \in \mathcal{F}$ be a Bayes decision function, with $(\bar{\bm{A}}, \bar{\bm{B}}, \bar{\bm{a}}, \bar{\bm{b}}, \bar{\bm{q}})$, and the number of latent factors be $\bar{K}$.  Let the truncated $V$-loss be $V_B(u) = \min(V(u), B)$, where $B \geq V\big(Y^t_{ij} \bar{f}^{t't}_{ij}(\bm{x}_{ij})\big)$ for any $i = 1, \cdots, n; j=1, \cdots, m; 0 \leq t' < t \leq T$, and the corresponding loss defined by $V_B(u)$ be 
\begin{eqnarray*}
L_{B}(\bff_{ij}, \bm{Z}_{ij}) = \sum_{0 \leq t'< t \leq T} w_{t' t} \Delta_{ij}  V_B \big(Y^t_{ij} f^{t't}_{ij}(\bm{X}_{ij}) \big) \mathbb{I}(Y^{t'}_{ij} = 1),
\end{eqnarray*}
and the truncated cost function is $l_{B}(\bff) = (nm)^{-1} \sum_{i,j} \mathbb{E}_{ij} L_{B}(\bff_{ij}, \bm{Z}_{ij})$. The following technical assumptions are assumed:


\noindent \textbf{Assumption A (Conversion)}. There exist constants $0 \leq \alpha \leq \infty$ and $c_1 >0$ 
such that for all $0 < \epsilon \leq B$ and $\bff \in \mathcal{F}$,
$$
\sup_{\{\bff \in \mathcal{F}; \ l_{B}(\bff) - l_{B}(\bar{\bff}) \leq \epsilon \}}  e(\bff) - e(\bar{\bff}) \leq c_1 \epsilon^\alpha.
$$

\noindent \textbf{Assumption B (Variance property)}. There exist constants $0 \leq \mu \leq 2$ and $c_2 > 0$ such that 
for any $\epsilon \geq 0$, $n, m \geq 1$, and $\bff \in \mathcal{F}$,
$$
\sup_{ \{\bff \in \mathcal{F}: \ l_{B}(\mathbf{f}) - l_{B}(\bar{\mathbf{f}}) \leq \epsilon \}} (nm)^{-1} \sum_{i=1}^n \sum_{j=1}^m \mathbb{E}_{ij} \Big( \big( L_{B}(\bff_{ij}, \bm{Z}_{ij}) - L_{B}(\bar{\bff}_{ij}, \bm{Z}_{ij}) \big)^2 \Big) \leq c_2 \epsilon^\mu.
$$

Assumptions A and B are two local smooth conditions regarding a connection between two different metrics and the relation between the mean and variance of the regret function in a neighborhood of $\bar{\mathbf{f}}$. Similar assumptions have been used in binary classification can be founded in \cite{zhang2014multicategory}, for example,  $\alpha = \mu = 1$ when $V(\cdot)$ is the $\psi$-loss \cite{shen2003psi} or the hinge loss under low-noise assumption.

Let $K$ and $\bar{K}$ be the numbers of latent factor of $\hat{\mathbf{f}}$ and $\bar{\mathbf{f}}$, respectively. 

\begin{theorem}
\label{theorem1}
Let $\hat{\bff}$ be a global minimizer of \eqref{model} with $K \geq \bar{K}$.  If Assumptions A and B hold, then there exist constants $c_3>0$ and $c_4>0$ such that
\begin{equation*}
\mathbb{P} \big( e(\hat{\mathbf{f}}) - e(\bar{\mathbf{f}}) \geq \epsilon^{2\alpha}_{nm} \big) \leq c_3 \exp \big(- c_4 nm(\lambda \bar{J} )^{2-\min(1,\mu)} \big),
\end{equation*}
where $\bar{J} = \max(J(\bar{\bff}),1)$, $J_\lambda(\bar{\bff}) \sim \epsilon^2_{nm}/2$, and  
\begin{equation}
\label{thm:rate}
\epsilon^2_{nm} = O\Big\{ \Big(\frac{(\Lambda+T)K}{nm} \log\big(  \frac{nm\bar{w}B\bar{J}(p_1 + p_2 + d_1 + d_2)(6T^3 + T^2 K)}{(\Lambda+T)K}  \big)\Big)^{\frac{1}{2-\min(1,\mu)}} \Big\},
\end{equation}
where $\bar{w} = \sum_{0 \leq t' < t \leq T} w^2_{t't}$, $(\Lambda + T)K$ is the total number of model parameters, and $\Lambda =p_1 + p_2 + \sum_{l=1}^{d_1} n_l + \sum_{l=1}^{d_2} m_l $.
\end{theorem}
When $nm \geq (p_1+p_2+d_1+d_2)(\Lambda+T)K(6T^3 + T^2 K)$, \eqref{thm:rate} suggests that the convergence rate for the proposed method is $O\big( \big(\frac{(\Lambda+T)K}{nm}\log(\frac{nm}{(\Lambda+T)K})\big)^{\frac{\alpha}{2 - \min(1,\mu)}}\big)$, which becomes $O\big(\frac{(\Lambda+T)K}{nm}\log(\frac{nm}{(\Lambda+T)K})\big)$ when $\alpha = \mu = 1$, and matches with the minimax rate of the standard classification in \citep{tsybakov2004optimal}. 

As indicated in \eqref{thm:rate}, the convergence rate is improved by the two-level monotonic chain property in \eqref{eqn:two_side} and the additive property in \eqref{eqn:additive}. As illustrated in Table \ref{tab:loopup}, based on the factorization model in \eqref{eqn:decision_function} while ignoring the two-level monotonicity \eqref{eqn:two_side}, a standard method and an ordinal method involve $\Lambda KT(T+1)/2$ and $\Lambda KT$ parameters, which are more than $(\Lambda + T)K$ and significantly impedes the performance when $T$ becomes large. Furthermore, due to the monotonic property in \eqref{eqn:property}, the effective sample size for $Y^{t'}_{ij} = 1$ decreases exponentially as $t'$ increases. Then the standard and ordinal methods suffer from over-fitting at late stages with a small sample size. In contrast, the proposed method leverages the two-level monotonicity to reduce the dimension of the underlying problem.

\begin{table*}[!ht]\centering
\scalebox{.8}{
\begin{tabular}{@{}llccccccllll@{}}  \toprule
 \phantom{a} & Method & \phantom{a} & \multicolumn{1}{c}{Forward monotonicity} & \phantom{a} & \multicolumn{1}{c}{Backward monotonicity}& \phantom{a} & \multicolumn{1}{c}{Consistency} & \phantom{a} & \multicolumn{1}{c}{\#Parameters} & \phantom{a}  \\ \midrule 
 & Proposed && \cmark && \cmark && \cmark && $(\Lambda + T) K$ \\
 & Standard && \xmark && \xmark && \xmark && $\Lambda KT(T+1)/2$ \\
 & Ordinal && \cmark && \xmark && \xmark && $\Lambda KT$ \\
\bottomrule
\end{tabular}
}
\caption{Monotonicity and model parameters of the proposed, standard, and ordinal frameworks, denoting \eqref{model}, \eqref{eqn:indv} and \eqref{eqn:ord}. Here forward and backward monotonicity is defined in \eqref{eqn:two_side}, consistency denotes if the methods can provide a consistent prediction result, and \#Parameters denotes the number of model parameters
based on \eqref{eqn:additive}.} 
\label{tab:loopup}
\end{table*}

\section{Numerical examples}
This section examines the proposed method in \eqref{model} with the hinge loss $V(u) = (1 - u)_+$ and compares it with standard methods, including latent factor models (SVD$^{++}$; \citep{koren2008factorization}), gradient boosting (GradBoost) \citep{friedman2001greedy}, support vector machine (SVM) \citep{cortes1995support}, deep neural network (DeepNN) \citep{schmidhuber2015deep}, ordinal method \eqref{eqn:ord}, namely ordinal support vector machine (OSVM), in a simulation and an article sharing benchmark.

For a standard method, SVD$^{++}$, GradBoost, SVM, and DeepNN treat each $(t', t)$ pair separately for all possible pairs, where we use one-hot encoder \citep{weinberger2009feature} to convert a categorical covariate to 0-1 numerical predictors for training. 
Moreover, for SVD$^{++}$, the number of latent factors is set as 20, and latent factors are estimated using the first columns of $\bm{s}_i$ and $\bm{o}_j$ with the rating $(Y^t_{ij}+1)/2 \in \{0, 1\}$ while the label is predicted as 1 if the predicted rating exceeds 0.5 and -1 otherwise; for GradBoost, the minimum number of samples to split an internal node is set to be 2 and  the minimum number of samples required to be at a leaf node is 1, and the number of boosting stages is set to be 10. For DeepNN, a ReLU network is used with the number of node in each layer being fixed at 32. For the ordinal method, we set the OSVM classifier in \eqref{eqn:ord} as $\mathbf{f}^{t't}_{ij} = (\bm{\beta}_{t'0} - \sum_{t_0=t'+1}^{t}\bm{\beta}_{t't_0})^\top \bm{x}_{ij}$ since $\bm{x}_{ij}$ is pre-scaled in $[0,1]$, where the forward monotonicity is ensured by positivity constraints $\bm{\beta}_{t't} \geq \bm{0}$, $t = t'+1, \cdots, T$.

For implementation, we use our Python library \texttt{varsvm} for the proposed method and OSVM, the Python library \texttt{sklearn}\footnote{\url{https://scikit-learn.org}} for GradBoost, SVM, and DeepNN, and the Python library \texttt{Surprise}\footnote{\url{http://surpriselib.com/}} for SVD$^{++}$. For all methods, the prediction of $Y_{ij}^{t}$ is automatically set as $-1$ when $Y^{t'}_{ij} = -1$.

\subsection{Simulations}

In simulations, data $\{\bm{s}_i, \bm{o}_j, y^t_{ij} \}_{i=1,\cdots, n; j=1, \cdots, m; t=0,\cdots, T}$ are generated to mimic a multistage recommender system on a monotonic chain, with $d_1=3$, $d_2=2$, $n=n_1=50$, $n_2=30$, $n_3 =50$, $m = m_1 = 100$, and $m_2 = 40$.  Then we set $K=20$, and for each $(l, h, t)$, $\bm{a}_{lh} \sim \chi^2(1)$, $\bm{b}_{lh} \sim \chi^2(1) $, and $\bm{q}_t \sim \chi^2(1)$ are independently sampled from the chi-squared distribution on 1 degree of freedom, and $(s_{il})_{l=1}^3$ and $(o_{jl})_{l=1}^2$ are uniformly sampled from $\{1, \cdots, n_l\}$ and $\{1, \dots, m_l\}$, respectively.
Moreover, we set $y^0_{ij} = 1$ at stage 0, set $p^1_{ij} = \bm{a}^\top(\bm{s}_i) \bm{b}(\bm{o}_j) - \big(\bm{a}(\bm{s}_i) \circ \bm{b}(\bm{o}_j)\big)^\top \bm{q}_1$, $y^1_{ij} = \sign \big(p^1_{ij} + \sigma_1 N(0,.1)\big)$ at stage 1, and set $p^2_{ij} = \bm{a}^\top(\bm{s}_i) \bm{b}(\bm{t}_j) - \big(\bm{a}(\bm{s}_i) \circ \bm{b}(\bm{t}_j)\big)^\top \bm{q}_2$, $y^2_{ij} = \sign \big(p^2_{ij} + \sigma_2 N(0,.1)\big)$ for $y^1_{ij} = 1$, and $y^2_{ij} = -1$ for $y^1_{ij} = -1$ at stage 2 to satisfy the monotonic chain property \eqref{eqn:property}, where $\sigma_t$ is the standard deviation for $(p^t_{ij})_{(i,j) \in \Omega_0}$; $t=1,2$. To generate dataset, we choose indices $(i, j)$ of  $\Omega_0$ randomly with $|\Omega_0| = 50000$, and the resulting missing ratio is 99.98\%, and $\Omega_1$ and $\Omega_2$ are generated as $\Omega_1 = \{ (i,j) \in \Omega_0: y^1_{ij} = 1\}$, $\Omega_2 = \{ (i,j) \in \Omega_0: y^2_{ij} = 1\}$. This leads to $\{\bm{s}_i, \bm{o}_j, \bm{y}_{ij} \}_{(i,j) \in \Omega_{0}}$. Finally, the data is split into training, validation, and testing sets with a partition ratio of 10\%, 10\%, and 80\%, respectively.

To avoid overfitting, for gradient boosting, the percentage of features are tuned via a grid search based on the overall misclassification error on the validation set, where the grid set is chosen as $\{.001, .005, .2, .4, .6, .8, 1. \}$.  For the neural network, its depth is tuned over a set $\{1, 2, 3, 5, 7, 9, 11\}$. For SVM, OSVM, and SVD$^{++}$, the penalization parameters are tuned based on validation set through $\{.001, .005, .01, .05, .1, .5, 1, 5, 10, 50, 100, 500\}$. For the proposed method, $K = 20$ and three tuning parameters $(\lambda_1, \lambda_2, \lambda_3)$ are tuned by minimizing the misclassification error on a validation set via a grid search over $\lambda_1 \in \{.001, .005, .01 \}$, $\lambda_2 \in \{ .01, .05, .1 \}$, and $\lambda_3 \in \{ .0001, .0005, .001 \}$. For evaluation, we set all stage weights $w_{t't} = 1$.

As suggested by Table \ref{tab:sim}, the proposed multistage recommender performs the best for one-step, two-step, and three-step forecasting, followed by DeepNN, SVM, and GradBoost. In terms of  the overall performance,  this same phenomenon is also observed and the amount of improvements of the proposed method are $43\%$, 22\%, $20\%$, and $13\%$ over Gradient Boosting, OSVM, SVM, and DeepNN. 
Concerning inconsistency prediction, SVM and OSVM perform the worst, followed by DeepNN and Gradient Boosting. 
Note that there is no single instance in which prediction by the proposed method is inconsistent, which is ensured by the two-level monotonicity property of the classier.   

\begin{table*}[!ht]\centering
  \scalebox{.69}{
  \begin{tabular}{@{}lcccccccccccccc@{}} \toprule
   (observe $t'$, predict $t$) & \phantom{a} & \multicolumn{1}{c}{Proposed} & \phantom{a} & \multicolumn{1}{c}{SVD$^{++}$}& \phantom{a}& \multicolumn{1}{c}{GradBoost} & \phantom{a}& \multicolumn{1}{c}{SVM} & \phantom{a}& \multicolumn{1}{c}{DeepNN} & \phantom{a}& \multicolumn{1}{c}{OSVM}  \\ \midrule 
   (Stage 0, Stage 1) && \textbf{.043(.000)} && .095(.001) && .062(.000) && .044(.000) && .044(.000) && .059(.000) \\
   (Stage 0, Stage 2) && \textbf{.120(.001)} && .387(.002) && .217(.000) && .153(.000) && .146(.000) && .165(.000) \\
   (Stage 0, Stage 3) && \textbf{.148(.000)} && .417(.003) && .272(.001) && .187(.000) && .185(.000) &&  .199(.000) \\
   (Stage 1, Stage 2) && \textbf{.097(.001)} && .294(.002) && .158(.000) && .103(.000) && .100(.000) && .116(.000) \\
   (Stage 1, Stage 3) && \textbf{.133(.000)} && .414(.002) && .246(.001) && .163(.000) && .160(.000) && .175(.000) \\
   (Stage 2, Stage 3) && \textbf{.095(.001)} && .341(.003) && .157(.001) && .101(.000) && .099(.000) && .102(.000) \\
   \midrule
   Overall && \textbf{.106(.000)} && .325(.001) && .185(.000) && .125(.000) && .122(.000) && .136(.000) \\
   \midrule
   \%Inconsist && \textbf{.000\%(.000)} && .680\%(.001) && 2.22\%(.000) && 10.9\%(.000) && 6.01\%(.000) && 10.9\%(.001) \\
  \bottomrule
  \end{tabular}}
  \caption{
  Test errors of six competitors and their estimated standard deviations in parentheses on simulated examples over 50 replications as well as a proportion of inconsistent instances violating monotonicity \eqref{eqn:two_side},denoted by \%Inconsist. Here proposed, SVD$^{++}$, GradBoost, SVM, DeepNN and OSVM denote the proposed method in \eqref{model} with the hinge loss, the latent factor model \citep{koren2008factorization}, gradient boosting \citep{friedman2001greedy}, support vector machine \citep{cortes1995support}, deep learner \citep{schmidhuber2015deep}, and conditional ordinal support vector machine in \eqref{eqn:ord}. The best performer in each case is bold-faced.}
  \label{tab:sim}
  \end{table*}


\subsection{Benchmark: \textit{Deskdrop} article sharing}


A \textit{Deskdrop} article sharing dataset  contains a sample of 12 months logs and about 73,000 logged user interactions on more than 3,000 public articles.  In particular, article-specific features are ``article Id'', ``author Id'', ``plain text'', as well as ``language'', and user-specific features include: ``user Id'', ``user agent'', and ``user region'', and users' actions are logged, including \textit{view}, \textit{like}, and \textit{follow}. The goal is to predict user-item interaction over three stage pairs.

For the proposed method, $\bm{v}_j$ is the numerical embedding based on Doc2Vec \citep{le2014distributed} of the ``plain text'' of item $j$, $\bm{o}_j$ consists of ``article Id'', ``author Id'', and ``language'' of item $j$, and $\bm{s}_i$ is composed of ``user Id'', ``user agent'', and ``user region'' of user $i$. For other competing methods, we use all features based on a one-hot encoder to convert categorical covariates to zero-one dummy predictors. Moreover, we predict $(t'=0, t = 1)$ based on the proposed method with $T=1$, and predict $(t'=0, t=2)$ and $(t'=1, t=2)$ based on the proposed method with $T=2$.

For evaluation, we set all stage weights $w_{t't} = 1$ and adjust the class weights at each stage inversely proportional to the number of observations in each class since feedback at each stage is highly imbalanced for this data. Hence, all methods are fitted based on balanced class weights or over- or under-sampling, except SVD$^{++}$ which based on a regression approach. Yet, for consistency, we focus on balanced zero-one loss for comparison since all methods are fitted and tuned based on a balanced-weighted classification loss.

The misclassification errors of all competing methods for each $(t',t)$ pair are computed based on a nested 15-fold cross-validation with an outer 5-fold loop and an inner 3-fold loop \citep{cawley2010over}. For the proposed method, $K = 7$ and three tuning parameters $(\lambda_1, \lambda_2, \lambda_3)$ are tuned by minimizing the misclassification error on a validation set via a grid search over $\lambda_1 \in \{.3, .5, .7, 1 \}$, $\lambda_2 \in \{ 2., 2.5, 3. \}$, and $\lambda_3 \in \{ .001, .005 \}$. For the neural network, SVM, OSVM, and SVD$^{++}$, the tuning parameters are selected with the same manner as simulation.

As suggested by Table \ref{tab:app}, the proposed multistage recommender outperforms SVD$^{++}$, GradBoost, SVM, DeepNN, and OSVM, in terms of the overall performance with the amount of improvement ranging from 75.3\% to 8.20\%, in terms of the class-balanced zero-one loss. Similarly, for one-step and two-step forward prediction, it is either the best or nearly the best. Concerning inconsistency prediction, all other methods have inconsistent cases ranging from 13.7\% to 0.98\%. Interestingly, the conditional ordinal method (OSVM) performs similarly to the conditional individual method (SVM), indicating that it does not fully account for the monotonicity property.

\begin{table*}[!ht]\centering
  \scalebox{.7}{
  \begin{tabular}{@{}lcccccccccccccc@{}} \toprule
   (observe $t'$, predict $t$) && \multicolumn{1}{c}{Proposed} & \phantom{a} & \multicolumn{1}{c}{SVD$^{++}$}& \phantom{a}& \multicolumn{1}{c}{GradBoost} & \phantom{a}& \multicolumn{1}{c}{SVM} & \phantom{a}& \multicolumn{1}{c}{DeepNN} & \phantom{a}& \multicolumn{1}{c}{OSVM} \\ \midrule 
   (Stage 0, Stage 1) && \textbf{0.136(.001)} && 0.499(.003) && 0.166(.000) && 0.139(.000) && 0.154(.001) && 0.140(.000)  \\
   (Stage 0, Stage 2) && \textbf{0.191(.001)} && 0.499(.007) && 0.218(.001) && 0.218(.001) && 0.223(.006) && 0.216(.001) \\
   (Stage 1, Stage 2) && \textbf{0.043(.000)} && 0.497(.007) && 0.049(.000) && 0.045(.000) && 0.139(.011) && 0.045(.000) \\
   \midrule
   Overall && \textbf{0.123(.001)} && 0.498(.006) && 0.144(.000) && 0.134(.000) && 0.172(.005) && 0.134(.000) \\
   \midrule 
   \%Inconsist &&  0.00\%(.000) && 7.79\%(.002) && 1.40\%(.000) && 1.90\%(.000) && 13.7\%(.005) && 0.98\%(.002) \\
  \bottomrule
  \end{tabular}
  }
  \caption{Class-balanced zero-one losses (CB-01) of six competitors and their estimated standard deviations in parentheses on the \textit{Deskdrop} benchmark, in addition to \%Inconsist denoting a proportion of inconsistent instances violating the monotonicity \eqref{eqn:two_side}. Here proposed, SVD$^{++}$, GradBoost, SVM, DeepNN and OSVM denote the proposed method in \eqref{model} with the hinge loss, the latent factor model \citep{koren2008factorization}, gradient boosting \citep{friedman2001greedy}, support vector machine \citep{cortes1995support}, deep learner \citep{schmidhuber2015deep}, and  ordinal support vector machine \eqref{eqn:ord}. The best performance in each case is bold-faced.}
  \label{tab:app}
\end{table*}

\newpage

\appendix
\section*{Technical Proofs}

\noindent \textbf{Proof of Lemma \ref{lem:bayes-rule}.} 
It suffices to show that $\bar{\bm{\phi}}^{t't}_{ij}$ is a global minimizer of $l_{ij}(\mathbf{f}_{ij})$. By \eqref{eqn:property}, 
\begin{align}
\label{pf:lemma1}
& \mathbb{E}_{ij}\big(\mathbb{I}(Y^t_{ij} {\phi}^{t't}_{ij}(\bm{x}_{ij}, Y^{t'}_{ij}) \leq 0 ) | \Delta_{ij} = 1 \big) \nonumber \\
& = \mathbb{P}_{ij}( Y^{t'}_{ij} = 1| \Delta_{ij} = 1) \mathbb{E}_{ij}\big(\mathbb{I}(Y^t_{ij} \phi^{t't}_{ij}(\bm{x}_{ij}, 1) \leq 0 ) | \Delta_{ij} = 1, Y^{t'}_{ij} = 1 \big) \nonumber \\
& \quad + \mathbb{P}_{ij}\big( Y^{t'}_{ij} = -1 | \Delta_{ij} = 1 \big) \mathbb{E}_{ij} \big( \mathbb{I}(-\phi^{t't}_{ij}(\bm{x}_{ij}, -1) \leq 0 ) | \Delta_{ij}=1, Y^{t'}_{ij} = -1 \big),
\end{align}
which yields that $\bar{\phi}^{t't}_{ij}(\bm{x}_{ij}, y^t_{ij})$ is a global minimizer of \eqref{pf:lemma1}, since $\bar{\phi}^{t't}_{ij}(\bm{x}_{ij},1) = \sign \big(\mathbb{P}_{ij}(Y^{t}_{ij} = 1 | Y^{t'}_{ij} = 1, \Delta_{ij}=1) - 1/2 \big)$ minimizes $\mathbb{E}_{ij}\big(\mathbb{I}(Y^t_{ij} {\phi}^{t't}_{ij}(\bm{x}_{ij}, 1) \leq 0 ) | Y^{t'}_{ij}=1, \Delta_{ij} = 1 \big)$ and $\bar{\phi}^{t't}_{ij}(\bm{x}_{ij}, -1) = -1$ minimizes $\mathbb{E}_{ij}\big( \mathbb{I}(-\phi_{ij}^{t't}(\bm{x}_{ij}, -1) \leq 0)|\Delta_{ij}=1, Y^{t'}_{ij} = -1 \big)$. 

Next, we verify the two-level monotonic property \eqref{eqn:two_side} of $\bar{f}^{t't}_{ij}$. Note that
\begin{align}
\mathbb{P}_{ij}(Y^t_{ij} = 1 | \Delta_{ij} = 1) & = \mathbb{P}_{ij}(Y^t_{ij} = 1, Y^{t-1}_{ij} = 1 | \Delta_{ij}=1) + \mathbb{P}_{ij}(Y^t_{ij} = 1, Y^{t-1}_{ij} = -1 | \Delta_{ij}=1) \nonumber \\
& = \mathbb{P}_{ij}(Y^t_{ij} = 1 | Y^{t-1}_{ij} = 1,  \Delta_{ij}=1) \mathbb{P}_{ij} ( Y^{t-1}_{ij} = 1 | \Delta_{ij} = 1) \nonumber \\
& \leq \mathbb{P}_{ij} ( Y^{t-1}_{ij} = 1 | \Delta_{ij} = 1),
\label{pf:monotonicity}
\end{align}
which implies that $\mathbb{P}_{ij}(Y^t_{ij} = 1 | \Delta_{ij} = 1)$ is nondecreasing over stage $t$. Moreover,
\begin{align*}
	& \mathbb{P}_{ij}(Y^t_{ij} = 1 | Y^{t'}_{ij} = 1, \Delta_{ij}=1) = \frac{\mathbb{P}_{ij}(Y^t_{ij} = 1, Y^{t'}_{ij} = 1 | \Delta_{ij}=1)}{\mathbb{P}_{ij}(Y^{t'}_{ij} = 1 | \Delta_{ij}=1)} = \frac{\mathbb{P}_{ij}(Y^t_{ij} = 1 | \Delta_{ij}=1)}{\mathbb{P}_{ij}(Y^{t'}_{ij} = 1 | \Delta_{ij}=1)}, \nonumber
\end{align*}
provided that $\mathbb{P}_{ij}(Y^{t'}_{ij} = 1 | \Delta_{ij}=1) \neq 0$, 
which is non-decreasing when $t$ increases while $t'$ is fixed because $\mathbb{P}_{ij}(Y^{t}_{ij} = 1 | \Delta_{ij}=1)$
is non-decreasing in $t$, and is non-increasing when $t'$ increases while $t$ is fixed because 
$\frac{1}{\mathbb{P}_{ij}(Y^{t'}_{ij} = 1 | \Delta_{ij}=1)}$ is non-increasing in $t'$. This implies
the two-level monotonic property.

Next, we construct an alternative form of the Bayes decision function $\bar{f}_{ij}^{t't}(\bm{x}_{ij})$, which
is additive in stage $t$:
$$
\bar{f}^{t't}_{ij}(\bm{x}_{ij}) = \bar{h}^0_{ij}(\bm{x}_{ij}) - \sum_{r = t'+1}^t \bar{h}^r_{ij}(\bm{x}_{ij}),
$$ 
where $\bar{h}^0_{ij}(\bm{x}_{ij}) = c_{ij}(\bm{x}_{ij}) \log_{\alpha_{ij}(\bm{x}_{ij})}(2) > 0$, $\bar{h}^r_{ij}(\bm{x}_{ij}) = c_{ij}(\bm{x}_{ij}) \log_{\alpha_{ij}(\bm{x}_{ij})} \big( \mathbb{P}_{ij}(Y_{ij}^{t-1} = 1 | \Delta_{ij} = 1) / \mathbb{P}_{ij}(Y_{ij}^{t}=1 | \Delta_{ij} = 1) \big) > 0$, and $c_{ij}(\bm{x}_{ij})>0$ and $ \alpha_{ij}(\bm{x}_{ij}) > 0$ are two arbitrary 
positive functions. Clearly, $\bar{\bar{f}}^{t't}_{ij}$ automatically satisfies the two-level monotonic property. 
Now, we need to
prove that $\bar{\bar{f}}^{t't}_{ij}(\bm{x}_{ij})$ has the same sign of $\bar{f}_{ij}^{t't}(\bm{x}_{ij})$, 
that is, $\sign\big(\bar\bar{f}^{t't}_{ij}(\bm{x}_{ij}) \big) = \sign\big(c_{ij}(\bm{x}_{ij}) \log_{\alpha_{ij}(\bm{x}_{ij})}\big( 2 \mathbb{P}_{ij}(Y^t_{ij} = 1 | Y_{ij}^{t'} = 1, \Delta_{ij} = 1) \big) \big) = 1$, when $\mathbb{P}_{ij}(Y^t_{ij} = 1 | Y_{ij}^{t'} = 1, \Delta_{ij} = 1) > 1/2$ and equals to $-1$ otherwise. The desired result then follows. \EOP

\noindent \textbf{Proof of Lemma \ref{lem:FC}.} The result follows from the fact that $\bar{\mathbf{f}}_{ij}^{t't}$ is a global minimizer of $\mathbb{E}_{ij}\big( L(\mathbf{f}_{ij}, \bm{Z}_{ij}) \big)$, when $V(\cdot)$ is Fisher consistency for binary classification.

\noindent \textbf{Proof of Lemma \ref{lem:algo}.} Note that \eqref{opt:user_A}, \eqref{opt:user_indv}, \eqref{opt:item_B}, \eqref{opt:item_b}, and \eqref{opt:stage} are convex minimization problems. Then convergence of
Algorithm 1 follows from \citep{tseng2001convergence}.  \EOP

\noindent \textbf{Proof of Theorem \ref{theorem1}.} Our treatment of bounding $\mathbb{P}\big(e(\hat{\bm{\phi}}) - e(\bar{\bm{\phi}}) \geq \epsilon^2_{nm} \big)$ relies on a chain argument of empirical process over a suitable partition of $\mathcal{F}$ induced by $e(\hat{\bm{\phi}}) - e(\bar{\bm{\phi}})$, as in \citep{ben2019rank,wong1995probability}. For $u \geq 1$ and $v \geq 0$, define
$A_{uv}  = \{ \mathbf{f} \in \mathcal{F}: 2^{u-1} \epsilon^2_{nm} \leq l_B(\bff) - l_B(\bar{\bff}) \leq 2^u \epsilon^2_{nm}, 2^{v-1}\bar{J} \leq J(f) \leq 2^v\bar{J} \}$ and
$A_{u0}  = \{\mathbf{f} \in \mathcal{F}: 2^{u-1} \epsilon^2_{nm} \leq l_B(\bff) - l_B(\bar{\bff}) \leq 2^u \epsilon^2_{nm}, J(f) \leq \bar{J} \}$.
By Assumption A,
\begin{align}
\label{proof:bound}
\mathbb{P}\big(e(\hat{\bff}) - e(\bar{\bff}) & \geq c_2 \epsilon_{nm}^{2\alpha} \big) \leq \mathbb{P}\big(l_B(\hat{\bff}) - l_B(\bar{\bff}) \geq \epsilon_{nm}^{2} \big) \nonumber \\
& \leq \mathbb{P}\big( \sup_{\{\bff \in \mathcal{F}; l_B(\bff) - l_B(\bar{\bff}) \geq \epsilon^2_{nm}\}} (nm)^{-1} \sum_{i,j} \big( \tilde{L}(\bar{\bff}_{ij}, \bm{Z}_{ij}) - \tilde{L}(\bff_{ij}, \bm{Z}_{ij}) \big) \geq 0 \big) \nonumber \\
& \leq \mathbb{P}\big( \sup_{\cup_{u=1}^{\infty} \cup_{v=0}^{\infty} A_{uv}} (nm)^{-1} \sum_{i,j} \big( \tilde{L}_B(\bar{\bff}_{ij}, \bm{Z}_{ij}) - \tilde{L}_B(\bff_{ij}, \bm{Z}_{ij}) \big) \geq 0 \big) \nonumber \\
& \leq \sum_{u=1}^\infty \sum_{v=1}^{\infty} \mathbb{P}\Big( \sup_{\{\bff \in A_{uv}\}} (nm)^{-1} \sum_{i,j} g(\bff_{ij}, \bm{Z}_{ij}) \geq \delta_{uv} \Big) \nonumber \\
& \quad + \sum_{u=1}^\infty \mathbb{P}\Big( \sup_{\{\bff \in A_{u0}\}} (nm)^{-1} \sum_{i,j} g(\bff_{ij}, \bm{Z}_{ij}) \geq \delta_{u0} \Big) =  \sum_{u=1}^\infty \sum_{v=1}^{\infty} I_{uv} + \sum_{u=1}^\infty I_{u0},
\end{align}
where $g(\bff_{ij}, \bm{Z}_{ij}) = L_B(\bar{\bff}_{ij}, \bm{Z}_{ij}) - L_B(\bff_{ij}, \bm{Z}_{ij}) - \big(L_B(\bar{\bff}_{ij}) - L_B({\bff}_{ij}) \big)$, $\delta_{uv} = 2^{u-1}\epsilon^2_{nm} + \lambda (2^{v-1} - 1)\bar{J}$, $\delta_{u0} = 2^{u-2} \epsilon^2_{nm}$, $I_{uv} = \mathbb{P}\big( \sup_{\{\bff \in A_{uv}\}} (nm)^{-1} \sum_{i,j} g(\bff_{ij}, \bm{Z}_{ij}) \geq \delta_{uv} \big)$, and $\lambda J(\bff) = \lambda_1 \|\bm{\beta}\|_2^2 + \sum_{k=1}^K \lambda_k \big( \|\bm{a}^k\|_2^2 + \|\bm{b}^k \|_2^2 \big)$.

To apply Talagrand's inequality \citep{gine2006concentration} to $I_{uv}$, let
$$\sigma^2_{uv} = \sup_{\{ \bff \in A_{uv} \}}  (nm)^{-1} \sum_{i,j} \Var_{ij} \big( g(\bff_{ij}, \bm{Z}_{ij}) \big), \quad u \geq 1, v \geq 0.$$ Then
\begin{align}
\label{proof:bound_inv}
I_{uv} & \leq \mathbb{P}\Big( \sup_{\{\bff \in A_{uv}\}} (nm)^{-1} \sum_{i,j} g(\bff_{ij}, \bm{Z}_{ij}) \geq \delta_{uv} \Big) \nonumber \\
& \leq \mathbb{P}\Big( \sup_{\{\bff \in A_{uv}\}} (nm)^{-1} \big| \sum_{i,j} g(\bff_{ij}, \bm{Z}_{ij}) \big| -\mathbb{E} \sup_{\{\bff \in A_{uv}\}} (nm)^{-1} \big| \sum_{i,j} g(\bff_{ij}, \bm{Z}_{ij})\big| \leq  \delta_{uv} - 2 \mathbb{E}\big( \mathcal{R}_{uv} \big) \Big) \nonumber \\
& \leq a_1 \exp(- \frac{nm \delta^2_{uv}}{ 2a_1(2\sigma_{uv}^2 + 16B \mathbb{E}(\mathcal{R}_{uv})+ B\delta_{uv})}) \leq a_1 \exp(- \frac{nm \delta^2_{uv}}{ 2a_1( 4^\mu c_2 \delta^\mu_{uv} + 5B\delta_{uv})}) \nonumber \\
& \leq a_1 \exp(- \frac{nm \delta^{2-\min(1, \mu)}_{uv}}{2a_1 a_2}),
\end{align}
where $a_2 = 32 c_2B^{\mu - \min(1, \mu)} + 9B^{2-\min(1,\mu)}$ and the second inequality follows from the fact that
\begin{align}
& \mathbb{E} \Big( (nm)^{-1} \sup_{\{\bff \in A_{uv}\}}  \big| \sum_{i,j} \big( L_B(\bar{\bff}_{ij}, \bm{Z}_{ij}) - L_B(\bff_{ij}, \bm{Z}_{ij}) - \mathbb{E}\big( L_B(\bar{\bff}_{ij}, \bm{Z}_{ij}) - L_B(\bff_{ij}, \bm{Z}_{ij}) \big) \big) \big| \Big) \nonumber \\
& \leq \mathbb{E} \Big( (nm)^{-1} \sup_{\{\bff \in A_{uv}\}} \big| \mathbb{E} \big( \sum_{i,j} \big( L_B(\bar{\bff}_{ij}, \bm{Z}_{ij}) - L_B(\bff_{ij}, \bm{Z}_{ij}) - (L_B(\bar{\bff}_{ij}, \bm{Z}'_{ij}) - L_B(\bff_{ij}, \bm{Z}'_{ij})) \big| \bm{Z} \big) \big| \Big) \nonumber \\
& \leq \mathbb{E} \Big( (nm)^{-1} \sup_{\{\bff \in A_{uv}\}} \big| \sum_{i,j} \big( L_B(\bar{\bff}_{ij}, \bm{Z}_{ij}) - L_B(\bff_{ij}, \bm{Z}_{ij}) - (L_B(\bar{\bff}_{ij}, \bm{Z}'_{ij}) - L_B(\bff_{ij}, \bm{Z}'_{ij})) \big| \Big) \nonumber \\
& \leq \mathbb{E} \Big( (nm)^{-1} \sup_{\{\bff \in A_{uv}\}} \big| \sum_{i,j} \tau_{i,j} \big( L_B(\bar{\bff}_{ij}, \bm{Z}_{ij}) - L_B(\bff_{ij}, \bm{Z}_{ij}) - (L_B(\bar{\bff}_{ij}, \bm{Z}'_{ij}) - L_B(\bff_{ij}, \bm{Z}'_{ij})) \big| \Big) \nonumber \\
& \leq 2\mathbb{E} \Big( (nm)^{-1} \sup_{\{\bff \in A_{uv}\}} \big| \sum_{i,j} \tau_{i,j} \big( L_B(\bar{\bff}_{ij}, \bm{Z}_{ij}) - L_B(\bff_{ij}, \bm{Z}_{ij}) \big| \Big) \leq 2 \mathbb{E}(\mathcal{R}_{uv}), \nonumber
\end{align}
where $\bm{Z}'_{ij}$ is an independent copy of $\bm{Z}_{ij}$, $\mathcal{R}_{uv} = \mathbb{E}_{\tau} \Big( (nm)^{-1} \sup_{\{\bff \in A_{uv}\}} \big| \sum_{i,j} \tau_{i,j} \big( L_B(\bar{\bff}_{ij}, \bm{Z}_{ij}) - L_B(\bff_{ij}, \bm{Z}_{ij}) \big| \Big)$ is the Rademacher complexity of $A_{uv}$, and $\tau_{ij}$ are independent random variables drawn from the Rademacher distribution. The third inequality follows from the Talagrand's inequality and the last inequality follows from the fact that,
\begin{align}
\label{proof:std_bound}
\sigma^2_{uv} = \sup_{\{ \bff \in A_{uv} \}}  (nm)^{-1} \sum_{i,j} \text{Var}_{ij} \big( L_B(\bff_{ij}, \bm{Z}_{ij}) - L_B(\bar{\bff}_{ij}, \bm{Z}_{ij}) \big) \leq c_2 (2^u \epsilon^2_{nm})^\mu \leq 16c_2\delta_{uv}^\mu.
\end{align}
A combination of \eqref{proof:bound} and \eqref{proof:bound_inv} yield
\begin{align}
& \mathbb{P}\big( e(\hat{\bff}) - e(\bar{\bff}) \leq c_2 \epsilon^{2\alpha}_{n,m} \big) \leq \sum_{u=1}^{\infty} \sum_{v=1}^\infty I_{uv} + \sum_{u=1}^\infty I_{u0} \nonumber \\
& \leq \sum_{i=1}^\infty \sum_{v=1}^\infty a_1 \exp \big(-\frac{nm \delta^{2-\min(1,\mu)}_{uv}}{2a_1a_2} \big) + \sum_{u=1}^\infty a_1 \exp \big(-\frac{nm \delta^{2-\min(1,\mu)}_{u0}}{2a_1a_2} \big) \nonumber \\
& \leq 4a_1 \exp \big(-\frac{nm(\lambda \bar{J})^{2 - \min(1, \mu)}}{4a_1a_2} \big). \nonumber
\end{align}
The desired result follows immediately. \EOP

\begin{lemma}
\label{lem:entropy}
Let $A_{uv} = \{ \mathbf{f} \in \mathcal{F}: 2^{u-1} \epsilon^2_{nm} \leq l_B(\bff) - l_B(\bar{\bff}) \leq 2^u \epsilon^2_{nm}, 2^{v-1}\bar{J} \leq J(f) \leq 2^v\bar{J} \}$, $A_{u0} = \{ \mathbf{f} \in \mathcal{F}: 2^{u-1} \epsilon^2_{nm} \leq l_B(\bff) - l_B(\bar{\bff}) \leq 2^u \epsilon^2_{nm}, J(f) \leq \bar{J} \}$, and $\delta_{uv} = 2^{u-1}\epsilon^2_{nm} + \lambda (2^{v-1} - 1)\bar{J}$, and $\delta_{u0} = 2^{u-2} \epsilon^2_{nm}$, where 
$$\epsilon^2_{nm} \sim \Big(\frac{(\Lambda+T)K}{nm} \log\big(  \frac{nm \bar{w} B\bar{J}\kappa(6T^3 + T^2K)}{(\Lambda+T)K}  \big)\Big)^{\frac{1}{2-\min(1,\mu)}}.$$ 
Then
\begin{equation}
\label{proof:entropy_condition}
4\mathbb{E}(\mathcal{R}_{uv}) \leq \delta_{uv},
\end{equation}
where $\mathcal{R}_{uv} = \mathbb{E}_{\tau} \Big( (nm)^{-1} \sup_{\{\bff \in A_{uv}\}} \big| \sum_{i,j} \tau_{i,j} \big( L_B(\bar{\bff}_{ij}, \bm{Z}_{ij}) - L_B(\bff_{ij}, \bm{Z}_{ij}) \big| \Big)$ is the Rademacher complexity of $A_{uv}$.
\end{lemma}

\noindent \textbf{Proof of Lemma \ref{lem:entropy}.} To bound $\mathbb{E}(\mathcal{R}_{uv})$, we compute the metric entropy for 
$$\mathcal{L}_{uv} = \big\{ (nm)^{-1} \sum_{i,j} \big( L_B(\bar{\bff}_{ij}, \bm{Z}_{ij}) - L_B(\bff_{ij}, \bm{Z}_{ij}) \big) : \bff \in A_{uv} \big\}.$$ For any functions $\tilde{\bff}$ and $\bff$,
\begin{align}
\label{eqn:diff}
(nm)^{-1} \sum_{i,j} \big(L_B(\tilde{\bff}_{ij}, \bm{Z}_{ij}) - L_B(\bff_{ij}, \bm{Z}_{ij}) \big)^2 & = \Big( \sum_{t' < t} w_{t't} \big( V_B(Y^t_{ij} \tilde{f}^{t't}_{ij}(\bm{x}_{ij})) - V_B(Y^t_{ij} f^{t't}_{ij}(\bm{x}_{ij})) \big) \Big)^2 \nonumber \\
& \leq \bar{w} \Big( \sum_{t' < t} \big( \tilde{f}^{t't}_{ij}(\bm{x}_{ij}) - f^{t't}_{ij}(\bm{x}_{ij}) \big)^2  \Big), 
\end{align}
where $\bar{w} = \sum_{0 \leq t' < t \leq T} w^2_{t't}$, and the first inequality follows from the Cauchy-Schwarz inequality. Next, we turn to bound $\tilde{f}^{t't}_{ij}(\bm{x}_{ij}) - f^{t't}_{ij}(\bm{x}_{ij})$,
\begin{align*}
& \big( \tilde{f}^{t't}_{ij}(\bm{x}_{ij}) - f^{t't}_{ij}(\bm{x}_{ij}) \big)^2 = \Big( \big( \tilde{a}(\bm{u}_i, \bm{s}_i) \circ \tilde{b}(\bm{v}_j, \bm{o}_j)\big)^\top \tilde{q}(t',t) - \big( a(\bm{u}_i, \bm{s}_i) \circ b(\bm{v}_j, \bm{o}_j)\big)^\top q(t',t) \Big)^2 \nonumber \\
& \leq  2 \Big( \big( \tilde{a}(\bm{u}_i, \bm{s}_i) \circ \tilde{b}(\bm{v}_j, \bm{o}_j)\big)^\top \big( \sum_{r=t'+1}^t (\bm{q}_r - \tilde{\bm{q}}_r) \big)  \Big)^2 \nonumber \\
& \hspace{3cm} + 2\Big( \big( \tilde{a}(\bm{u}_i, \bm{s}_i) \circ \tilde{b}(\bm{v}_j, \bm{o}_j) - a(\bm{u}_i, \bm{s}_i) \circ b(\bm{v}_j, \bm{o}_j)\big)^\top q(t',t) \Big)^2 = I_1 + I_2,
\end{align*}
where $q(t', t) = \bm{1} - \sum_{r = t'+1}^t \bm{q}_r$. Then we bound $I_1$ and $I_2$ separately. For $I_1$,
\begin{align}
I_1 & \leq 2 \big\| \tilde{a}(\bm{u}_i, \bm{s}_i) \circ \tilde{b}(\bm{v}_j, \bm{o}_j) \big \|_2^2 \big \| \sum_{r=t'+1}^t \bm{q}_r - \tilde{\bm{q}}_r \big\|_2^2 \leq 2 \big\| \tilde{a}(\bm{u}_i, \bm{s}_i) \big\|_2^2 \big\| \tilde{b}(\bm{v}_j, \bm{o}_j) \big \|_2^2 \big \| \sum_{r=t'+1}^t \bm{q}_r - \tilde{\bm{q}}_r \big\|_2^2 \nonumber \\
& \leq 8(t - t') (p_1 + d_1) (p_2 + d_2) \big( \| \bm{A} \|_F^2 + \|\bm{a}\|_2^2 \big) \big( \| \bm{B} \|_F^2 + \|\bm{b}\|_2^2 \big) \big( \sum_{r=t'+1}^t \big \| \bm{q}_r - \tilde{\bm{q}}_r \big\|_2^2 \big) \nonumber \\
& \leq 2^{2v+3} \bar{J}^2 (t - t') (p_1 + d_1) (p_2 + d_2)   \big( \sum_{r=t'+1}^t \big \| \bm{q}_r - \tilde{\bm{q}}_r \big\|_2^2 \big),
\label{eqn:I1}
\end{align}
where the last inequality follows from the fact that
$$
\big\| a(\bm{u}_i, \bm{s}_i) \big\|_2^2 \leq 2 \big( \|\bm{A}\|_F^2 \|\bm{u}_i\|_2^2 + d_1 \sum_{l=1}^{d_1} \max_{i=1, \cdots, n_l} \|\bm{a}_{li}\|_2^2 \big) \leq 2 \big( p_1 \|\bm{A}\|_F^2 + d_1 \|\bm{a}\|_2^2 \big),
$$
and 
$$
\big\| b(\bm{v}_j, \bm{o}_j) \big\|_2^2 \leq 2 \big( \|\bm{B}\|_F^2 \| \bm{v}_j \|_2^2 + d_2 \sum_{l=1}^{d_2} \max_{j=1, \cdots, m_l} \|\bm{b}_{lj}\|_2^2 \big) \leq 2 \big( p_2 \|\bm{B}\|_F^2 + d_2 \|\bm{b}\|_2^2 \big).
$$
For $I_2$,
\begin{align}
I_2 & \leq 2\Big( \big( \tilde{a}(\bm{u}_i, \bm{s}_i) \circ \tilde{b}(\bm{v}_j, \bm{o}_j) - a(\bm{u}_i, \bm{s}_i) \circ b(\bm{v}_j, \bm{o}_j)\big)^\top q(t',t) \Big)^2 \nonumber \\
& \leq 2 \big\| \bm{1} - \sum_{r=t'+1}^t \bm{q}_r \big\|_2^2 \big\| \tilde{a}(\bm{u}_i, \bm{s}_i) \circ \tilde{b}(\bm{v}_j, \bm{o}_j) - a(\bm{u}_i, \bm{s}_i) \circ b(\bm{v}_j, \bm{o}_j) \big\|_2^2  \nonumber \\
& \leq 4 \big(K + (t-t')\sum_{r=t'+1}^t \|\bm{q}_r\|_2^2 \big) \Big\| \tilde{a}(\bm{u}_i, \bm{s}_i) \circ \tilde{b}(\bm{v}_j, \bm{o}_j) - \tilde{a}(\bm{u}_i, \bm{s}_i) \circ b(\bm{v}_j, \bm{o}_j) \Big\|_2^2 \nonumber \\
& \quad + 4 \big(K + (t-t')\sum_{r=t'+1}^t \|\bm{q}_r\|_2^2 \big) \Big\| \tilde{a}(\bm{u}_i, \bm{s}_i) \circ b(\bm{v}_j, \bm{o}_j) - a(\bm{u}_i, \bm{s}_i) \circ b(\bm{v}_j, \bm{o}_j) \Big\|_2^2 \nonumber \\
& \leq 4 \big(K + (t-t')\sum_{r=t'+1}^t \|\bm{q}_r\|_2^2 \big) \big\| \tilde{a}(\bm{u}_i, \bm{s}_i) \big\|_2^2 \big\| (\tilde{\bm{B}} - \bm{B})\bm{v}_j - \sum_{l=1}^{d_2} (\tilde{\bm{b}}_{lo_{jl}} - \bm{b}_{lo_{jl}}) \big\|_2^2 \nonumber \\
& \quad + 4\big(K + (t-t')\sum_{r=t'+1}^t \|\bm{q}_r\|_2^2 \big) \big\| b(\bm{v}_j, \bm{o}_j) \big\|_2^2 \big\| (\tilde{\bm{A}} - \bm{A})\bm{u}_i - \sum_{l=1}^{d_1} (\tilde{\bm{a}}_{ls_{il}} - \bm{a}_{ls_{il}}) \big\|_2^2 \nonumber \\
& \leq 2^{v+4} \kappa \bar{J} \big(K + (t-t')\sum_{r=t'+1}^t \|\bm{q}_r\|_2^2 \big)  \nonumber \\
& \hspace{4cm} \Big( p_1 \| \tilde{\bm{A}} - \bm{A} \|_F^2 + p_2 \| \tilde{\bm{B}} - \bm{B} \|_F^2 + d_1 \| \tilde{\bm{a}} - \bm{a} \|_2^2 + d_2 \| \tilde{\bm{b}} - \bm{b} \|_2^2 \Big),
\label{eqn:I2}
\end{align}
where $\kappa = p_1 + p_2 + d_1 + d_2$. Combining \eqref{eqn:diff}, \eqref{eqn:I1} and \eqref{eqn:I2},
\begin{align}
& (nm)^{-1} \sum_{i,j} \big(L_B(\tilde{\bff}_{ij}, \bm{Z}_{ij}) - L_B(\bff_{ij}, \bm{Z}_{ij}) \big)^2 \nonumber \\
& \leq \bar{w} \Big( 2^{2v+6} \bar{J}^2 (p_1 + d_1) (p_2 + d_2) \sum_{t' < t} (t - t') \sum_{r=t'+1}^t \big \| \bm{q}_r - \tilde{\bm{q}}_r \big\|_2^2  \nonumber \\
& \ + \kappa^2 2^{v+4} \bar{J} \big( \| \tilde{\bm{A}} - \bm{A} \|_F^2 + \| \tilde{\bm{B}} - \bm{B} \|_F^2 + \| \tilde{\bm{a}} - \bm{a} \|_2^2 + \| \tilde{\bm{b}} - \bm{b} \|_2^2 \big) \big( \sum_{t' < t} \big(K + (t-t')\sum_{r=t'+1}^t \|\bm{q}_r\|_2^2 \big) \big)  \Big) \nonumber \\
& \leq \bar{w} \Big( 2^{2v+2} \bar{J}^2 T^3 (p_1 + d_1) (p_2 + d_2)  \big\| \bm{q} - \tilde{\bm{q}} \big\|_2^2  \nonumber \\
& \ + \kappa^2 2^{v} \bar{J}  \big( (T+1)TK + T^3 \|\bm{q} \|_2^2 \big) \big( \| \tilde{\bm{A}} - \bm{A} \|_F^2 + \| \tilde{\bm{B}} - \bm{B} \|_F^2 + \| \tilde{\bm{a}} - \bm{a} \|_2^2 + \| \tilde{\bm{b}} - \bm{b} \|_2^2 \big)  \Big). \nonumber
\end{align}
Therefore, the metric entropy for $\mathcal{L}_{uv}$ for $\nu \leq 1$ is upper bounded:
\begin{align}
\log \mathcal{N}(\nu, \mathcal{L}_{uv}) & \leq \log \mathcal{N} \Big( \frac{\nu}{2^{v} \bar{J}Ba_3}, \mathcal{B}_+  \big((\Lambda+T)K\big) \Big) \leq \gamma \log \big( \frac{2^{v+1} \bar{J}Ba_3}{\nu} \big),
\end{align}
where $a_3 = \kappa  \sqrt{\bar{w}(6T^3 + (T+1)TK)}$, $\mathcal{B}_+((\Lambda+T)K)$ is the unit $l_2$-ball in $\mathbb{R}_+^{(\Lambda+T)K}$. Therefore, the Rademacher complexity $\mathbb{E}(\mathcal{R}_{uv})$ can be computed based on Theorem 3.12 of \citep{koltchinskii2011oracle}. Specifically, there exists a constant $a_4$, such that
$$
\mathbb{E}\big( \mathcal{R}_{uv} \big) \leq a_4 \max\Big( \frac{\delta^{\mu/2}_{uv}\sqrt{(\Lambda+T)K}}{\sqrt{nm}} \sqrt{\log \big( \frac{2^{v+1}\bar{J}Ba_3}{\delta^{\mu/2}_{uv}}\big)}, \frac{B(\Lambda+T)K}{nm} \log \big( \frac{2^{v+1}\bar{J}Ba_3}{\delta^{\mu/2}_{uv}}\big)  \Big).
$$
Then there exists a constant $a_5>0$, such that $\epsilon^2_{nm} = a_5 \Big(\frac{(\Lambda+T)K}{nm} \log\big(  \frac{nm\bar{w}B\bar{J}\kappa(6T^3 + T^2K)}{(\Lambda+T)K}  \big)\Big)^{\frac{1}{2-\min(1,\mu)}}$, and
\begin{align}
\frac{\delta_{uv}}{4\mathbb{E}\big( \mathcal{R}_{uv} \big)} & \geq \frac{1}{4a_4} \min \Big( \frac{\sqrt{nm}\delta^{1-\mu/2}_{uv} }{\sqrt{(\Lambda+T)K} } \big(\log \big( \frac{2^{v+1}\bar{J}Ba_3}{\delta^{\mu/2}_{uv}}\big)\big)^{-1/2}, \frac{nm \delta_{uv}}{B(\Lambda+T)K} \big(\log \big( \frac{2^{v+1}\bar{J}Ba_3}{\delta^{\mu/2}_{uv}}\big)\big)^{-1}  \Big) \nonumber \\
& \geq \frac{1}{4a_4} \min \Big( \frac{\sqrt{nm}\delta^{1-\mu/2}_{u0} }{\sqrt{(\Lambda+T)K} } \big(\log \big( \frac{2\bar{J}Ba_3}{\delta^{\mu/2}_{u0}}\big)\big)^{-1/2}, \frac{nm \delta_{u0}}{B(\Lambda+T)K} \big(\log \big( \frac{2\bar{J}Ba_3}{\delta^{\mu/2}_{u0}}\big)\big)^{-1} \Big) \nonumber \\
& \geq \frac{1}{4a_4} \min \Big( \frac{\sqrt{nm}\epsilon^{2-\mu}_{nm} }{\sqrt{(\Lambda+T)K} } \big(\log \big( \frac{2\bar{J}Ba_3}{\epsilon^{\mu}_{nm}}\big)\big)^{-1/2}, \frac{nm \epsilon^2_{nm}}{B(\Lambda+T)K} \big(\log \big( \frac{2\bar{J}Ba_3}{\epsilon^{\mu}_{nm}}\big)\big)^{-1}  \Big) \geq 1.
\nonumber
\end{align}
where the first two inequalities follow from non-decreasing of $(u,v)$. This completes the proof. \EOP


\vskip 0.2in
\bibliography{causal_rs}
\end{document}